\documentclass[twocolumn]{aastex7}
\usepackage{amsmath}

\newcommand{\vone}[1]{{#1}}
\newcommand{\vtwo}[1]{{#1}}

\begin{document}

\title{NASA’s Pandora SmallSat Mission: Gaussian Processes for Simulated Stellar Heterogeneity Corrections under Imperfect Stellar Models}

\author[0000-0002-4962-2543]{Zoe Ko}
\affiliation{Space Science Institute, Lawrence Livermore National Laboratory, 7000 East Ave., Livermore, CA 94550, USA}
\affiliation{William H. Miller III Department of Physics and Astronomy, Johns Hopkins University, Baltimore, MD 21218, USA}
\email{zko2@jh.edu}

\author[0000-0002-1052-6749]{Peter McGill}
\affiliation{Space Science Institute, Lawrence Livermore National Laboratory, 7000 East Ave., Livermore, CA 94550, USA}
\email{mcgill5@llnl.gov}

\author[0000-0002-3627-1676]{Benjamin V. Rackham}
\affiliation{Department of Earth, Atmospheric and Planetary Sciences, Massachusetts Institute of Technology, 77 Massachusetts Avenue, Cambridge, MA 02139, USA} 
\affiliation{Kavli Institute for Astrophysics and Space Research, Massachusetts Institute of Technology, Cambridge, MA 02139, USA}
\email{brackham@mit.edu}

\author[0000-0003-3714-5855]{D\'aniel Apai}
\affiliation{Steward Observatory, University of Arizona, 933 N. Cherry Avenue, Tucson, AZ 85721, USA} 
\affiliation{Lunar and Planetary Laboratory, University of Arizona, 1629 E. University Boulevard, Tucson, AZ 85721, USA}
\email{apai@arizona.edu}

\author[0000-0003-0971-1709]{Aishwarya R. Iyer}
\affiliation{NASA Goddard Space Flight Center, 8800 Greenbelt Road, Greenbelt, MD 20771, USA}
\affiliation{NASA Postdoctoral Fellowship through Oak Ridge Associated Universities}
\email{aishwarya.iyer@nasa.gov}

\author[0000-0002-5904-1865]{Jason F. Rowe}
\affiliation{Department of Physics and Astronomy, Bishop's University, 2600 Rue College, Sherbrooke, QC J1M 1Z7, Canada}
\email{jason.rowe@ubishops.ca}

\author[0000-0003-0156-4564]{Luis Welbanks}
\affiliation{School of Earth \& Space Exploration, Arizona State University, Tempe, AZ 85287, USA}
\email{luis.welbanks@asu.edu}

\author[0000-0001-8018-0264]{Suman Saha}
\affiliation{Steward Observatory, University of Arizona, 933 N. Cherry Avenue, Tucson, AZ 85721, USA}
\email{sumansaha@arizona.edu}

\author[0000-0003-4459-9054]{Yoav Rotman}
\affiliation{School of Earth \& Space Exploration, Arizona State University, Tempe, AZ 85287, USA}
\email{yrotman@asu.edu}


\author[0000-0003-1309-2904]{Elisa V. Quintana}	
\affiliation{NASA Goddard Space Flight Center, 8800 Greenbelt Road, Greenbelt, MD 20771, USA}
\email{elisa.quintana@nasa.gov}

\author[orcid=0000-0003-4206-5649]{Jessie L. Dotson}
\affiliation{NASA Ames Research Center, MS 245-6, Moffett Field, CA 94035, USA} \email{jessie.dotson@nasa.gov}

\author[0000-0001-8020-7121]{Knicole D. Col\'{o}n}
\affiliation{NASA Goddard Space Flight Center, 8800 Greenbelt Road, Greenbelt, MD 20771, USA}
\email{knicole.colon@nasa.gov}

\author[0000-0001-7139-2724]{Thomas Barclay} 
\affiliation{NASA Goddard Space Flight Center, 8800 Greenbelt Road, Greenbelt, MD 20771, USA} 
\email{thomas.barclay@nasa.gov}

\author[0000-0002-3385-8391]{Christina Hedges}
\email{christina.l.hedges@nasa.gov, chedges@umbc.edu}
\affiliation{University of Maryland, Baltimore County, 1000 Hilltop Circle, Baltimore, Maryland, United States}
\affiliation{NASA Goddard Space Flight Center, Greenbelt, Maryland, United States}

\author[0000-0002-0388-8004]{Emily A. Gilbert} 
\affiliation{Caltech/IPAC-NASA Exoplanet Science Institute, Pasadena, CA 91125, USA} 
\email{egilbert@caltech.edu}

\author[0000-0001-9513-1449]{N\'estor Espinoza}
\affiliation{Space Telescope Science Institute, 3700 San Martin Drive, Baltimore, MD 21218, USA}
\affiliation{William H. Miller III Department of Physics and Astronomy, Johns Hopkins University, Baltimore, MD 21218, USA}
\email{nespinoza@stsci.edu}

\author[0000-0002-6276-1361]{Trevor O. Foote} 
\affiliation{NASA Goddard Space Flight Center, 8800 Greenbelt Road, Greenbelt, MD 20771, USA} 
\affiliation{NASA Postdoctoral Fellowship through Oak Ridge Associated Universities} 
\email{trevor.o.foote@nasa.gov}

\author[0000-0002-5034-9476]{Rae Holcomb} 
\affiliation{NASA Goddard Space Flight Center, 8800 Greenbelt Road, Greenbelt, MD 20771, USA} 
\affiliation{NASA Postdoctoral Fellowship through Oak Ridge Associated Universities}
\email{raeholcomb15@gmail.com}

\author[0000-0001-5084-4269]{Benjamin J. Hord} 
\affiliation{NASA Goddard Space Flight Center, 8800 Greenbelt Road, Greenbelt, MD 20771, USA} 
\affiliation{NASA Postdoctoral Fellowship through Oak Ridge Associated Universities} 
\email{benjamin.j.hord@nasa.gov}

\author[0000-0003-2528-3409]{Brett M. Morris} 
\affiliation{Space Telescope Science Institute, 3700 San Martin Drive, Baltimore, MD 21218, USA}
\email{bmmorris@stsci.edu}

\author[0000-0002-8619-8542]{Joe P. Renaud}
\affiliation{University of Maryland, College Park, Maryland, United States} 
\affiliation{NASA Goddard Space Flight Center, 8800 Greenbelt Road, Greenbelt, MD 20771, USA}
\email{joseph.p.renaud@nasa.gov}


\begin{abstract}
\noindent Stellar contamination is the primary bottleneck limiting robust atmospheric characterization of exoplanets orbiting active K- and M-dwarfs with James Webb Space Telescope (JWST) transmission spectroscopy. Launched in January 2026, NASA's Pandora SmallSat mission \vone{addresses} this challenge through long-term, simultaneous visible photometry and near-infrared spectroscopy of planet-hosting stars, providing the data to disentangle planetary and stellar signals. However, conventional stellar-contamination corrections rely on stellar atmosphere models, making them vulnerable to model inaccuracies that can impede accurate inferences. Here, we investigate how Pandora data can be used to mitigate stellar contamination without an accurate host star model. We present a Gaussian-process-aided stellar retrieval framework that learns wavelength-dependent deviations from current stellar model grids. \vone{We propagate stellar contamination factors directly from the stellar retrieval into the planetary atmospheric retrievals,} allowing the atmospheric inference to marginalize over stellar-contamination uncertainty without relying on a biased stellar model. Tests using simulated Pandora observations with realistic stellar model mismatch \vone{show} that this framework consistently improves the recovery of disk-integrated stellar spectra. \vone{For a WASP-69 b-like system, this approach can reduce model-driven biases in inferred atmospheric properties, though degeneracies among stellar and atmospheric parameters can leave residual biases. This flexible, data-driven framework is a promising approach toward accurate stellar-contamination corrections and more robust atmospheric inference even when host-star model grids are inaccurate.}
\end{abstract}


\section{Introduction} 
JWST transmission spectroscopy routinely reaches precisions at the tens-of-ppm level \citep[e.g.][]{rustamkulov2023, lim2023, kirk2024, beatty2024}, making stellar contamination, or the transit light source effect, a leading limitation on exoplanet atmospheric characterization \citep{rackham2018a, rackham2018b, rackham2023, lim2023, moran2023, radica2025, espinoza2025trappist, ko2026}. By measuring starlight that filters through a planet's atmosphere, transmission spectroscopy is inevitably sensitive to the stellar photosphere. Heterogeneities like spots and faculae can imprint signatures onto transmission spectra, masking or mimicking exoplanet atmospheric features \citep{Rackham2017, rackham2018a, rackham2018b, pinhas2018, zhang2018, wakeford2019, garcia2022}. These effects are especially prominent for K- and M-dwarfs due to their high magnetic activity and large spot and facula covering fractions \citep{rackham2024, smitha2025, sumida2026, murray2026}. For several rocky planets around such hosts, stellar contamination has prevented definitive interpretations of the transmission spectra \citep[e.g.][]{moran2023, lim2023, may2023, barclay2025b, allen2026}. Developing models to address stellar contamination is therefore essential to robustly understanding exoplanet atmospheres.

Successfully launched in January 2026 \citep{Foote2026} as part of NASA’s Astrophysics Pioneers Program, the Pandora SmallSat mission aims to address the issue of stellar contamination by obtaining long-baseline observations of a sample of exoplanets and their host stars to disentangle planetary and stellar signals \citep{barclay2025a}. Pandora’s visible-band photometric channel (VISDA) covers 0.4 to 0.7~$\micron$ and is responsible for tracking the time-varying stellar surface heterogeneities \citep{karburn2023}. The near-infrared spectroscopic channel (NIRDA) covers 0.9 to 1.6~$\micron$ with an effective resolving power of $R \sim 120$ and is responsible for detecting exoplanet molecular features such as H$_2$O and CH$_4$ \citep{rotman2026}. The primary mission sample consists of around 20 planets spanning Earth to Jupiter in size, orbiting active K- or M-dwarfs. For each target, Pandora will obtain around ten 24-hour visits \citep{barclay2025a, Gilbert2026}. By operating concurrently with JWST, Pandora will also provide long-baseline stellar monitoring needed to correct stellar contamination in JWST transmission spectra.

Under the key assumption that the adopted stellar model perfectly describes the true stellar spectra, Pandora can typically recover photospheric temperatures to $\sim 30$\,K, infer the stellar radius with negligible bias at the $<0.1\%$ level, and constrain spot properties when they contribute noticeably to the disk-integrated spectrum. For simple spot geometries, Pandora’s out-of-transit data can reduce contamination signals from $\sim 10^2$–$10^3$~ppm down to $\lesssim 10$~ppm, well below the anticipated precision of Pandora transmission spectra \citep{rackham2026pand}.

However, stellar models are imperfect representations of real stellar spectra. Typical 1D spectral models neglect the magnetic and 3D convective structure of active regions, treating spot spectra as temperature shifts of the quiescent photosphere. Independent observational and theoretical work has shown that this approximation breaks down for cool stars, especially below 3000\,K \citep{freytag2010, wakeford2019, garcia2022, murray2026, iyer2026, iyer2020, narrett2024, smitha2025, sumida2026}. Therefore, the accuracy of contamination corrections is constrained by the fidelity of the underlying stellar models.

In this paper, we directly address the issue of stellar model--data mismatch by introducing two Gaussian process (GP) components into the stellar spectral model: one describing deviations in the photosphere spectrum and one describing deviations in the spot spectrum. GPs provide a flexible, non-parametric framework for modeling wavelength-correlated residuals and are thus able to capture and marginalize over stellar model inaccuracies without assuming a specific functional form. GPs have been explored as a way to model correlated residuals in transmission spectra \citep{rotman2025} and have been incorporated into JWST analyses to tackle stellar contamination \citep{espinoza2025trappist, ko2026}. However, individual JWST transit observations generally lack the long-baseline, time-variable stellar monitoring needed to empirically disentangle the underlying stellar spectral components. Pandora is well-suited to address this problem because its repeated out-of-transit observations can sample changing spot covering fractions as stars rotate over time, causing the photosphere and spot components to contribute with different weights to the observed disk-integrated spectrum. Through this time-variable mixing, Pandora's out-of-transit data can be used to separate the two components and infer the stellar contamination signal. This is analogous to existing GP-based spectral disentangling frameworks that separate binary-star components from time-variable composite spectra \citep{czekala2017} \vone{and methods that use time-resolved spectroscopy over a stellar rotation cycle to empirically constrain the spectra of stellar photospheres and surface heterogeneities \citep{berardo2024}. Our framework follows this empirical disentangling motivation but is tailored to Pandora's observing strategy and addresses stellar-model mismatch by using stellar models as GP mean functions rather than exact templates.} This directly supports Pandora’s science goal of measuring the stellar spot covering fractions and correcting their impacts on transmission spectra \citep{barclay2025a}. Using \vone{revised} simulated Pandora datasets \vone{based on} \citet{rackham2026pand}, we quantify how this GP-aided stellar model framework improves the recovery of latent stellar components and corrected transmission spectra relative to retrievals that rely solely on biased physical stellar models.

We also propagate the stellar retrieval posterior samples into the atmospheric retrieval as empirical prior distributions over the contamination correction factor. The atmospheric retrieval is thus able to incorporate stellar-contamination uncertainty directly into the inferred planetary parameters. This differs from conventional approaches that correct a transmission spectrum using deterministic stellar templates, which can lead to biased and overconfident atmospheric constraints when the stellar models are inaccurate.

In Section~\ref{sec:simulated}, we describe the simulated Pandora dataset used to test this framework. In Section~\ref{sec:retrieval-set-up}, Methods I, we present the stellar retrieval framework, including the baseline two-component stellar contamination model and the GP-aided extension. In Section~\ref{sec:transmission-propagation}, Methods II, we describe how the inferred stellar spectra are propagated into atmospheric retrievals. In Section~\ref{sec:results}, we present results for a representative M-dwarf scenario and then summarize performance across all eight simulated stellar activity scenarios. In Section~\ref{sec:discussion}, we interpret the GP corrections, discuss when the GP-aided framework is warranted, and outline future extensions. Finally, we summarize our conclusions in Section~\ref{sec:conclusion}.

\section{Simulated Dataset} \label{sec:simulated}

\vone{We use a revised set of simulated Pandora observations based on the eight stellar activity scenarios presented by \citet{rackham2026pand}. 
The spot filling-factor time series have been rescaled to produce approximately 1\% peak-to-peak VISDA variability across the sampled observations, representative of Pandora targets. 
This revised dataset is publicly available on Zenodo\footnote{DOI: \dataset[10.5281/zenodo.23990909]{\doi{10.5281/zenodo.23990909}}}.}
This dataset consists of simulated out-of-transit Pandora VISDA and NIRDA observations for eight representative stellar activity scenarios: K- and M-dwarfs, fast and slow rotators, giant and solar-like spots (e.g., MFG denotes an M-dwarf, fast rotator with giant spots). Each scenario is set to a magnitude of $J=9$~mag, representative of Pandora targets. The dataset is summarized in Table~\ref{tab:simulated-scenarios}.

\begin{table*}
    \centering
    \caption{\vone{Simulated stellar activity scenarios. $R_{\rm spot}$ is the spot angular radius, and $f_{\rm min}$ and $f_{\rm max}$ are the minimum and maximum spot covering fractions of the sinusoidal time series. The NIRDA and VISDA variabilities are peak-to-peak amplitudes as a percentage of the mean flux. Distances are set such that an unspotted star with the given photosphere temperature has $J=9$.}}
    \label{tab:simulated-scenarios}
    \scriptsize
    \begin{tabular}{lccccccccccc}
        \hline
        \hline
        Scenario & $T_{\rm phot}$ (K) & $T_{\rm spot}$ (K) & $P_{\rm rot}$ (d) & $R_\star$ ($R_\odot$) & $\log g$ & $D_\star$ (pc) & $R_{\rm spot}$ ($^\circ$) & $f_{\rm min}$ & $f_{\rm max}$ & NIRDA var. (\%) & VISDA var. (\%) \\
        \hline
        KFG & 4700 & 3600 & 5  & 0.73 & 4.5 & 69.2 & 7 & 0.000 & 0.011 & 0.63 & 1.00 \\
        KFS & 4700 & 3600 & 5  & 0.73 & 4.5 & 69.2 & 2 & 0.030 & 0.041 & 0.61 & 1.00 \\
        KSG & 4700 & 3600 & 30 & 0.73 & 4.5 & 69.2 & 7 & 0.000 & 0.011 & 0.62 & 1.00 \\
        KSS & 4700 & 3600 & 30 & 0.73 & 4.5 & 69.2 & 2 & 0.030 & 0.041 & 0.65 & 1.00 \\
        MFG & 3400 & 3000 & 5  & 0.36 & 5.0 & 21.1 & 7 & 0.000 & 0.013 & 0.50 & 1.00 \\
        MFS & 3400 & 3000 & 5  & 0.36 & 5.0 & 21.1 & 2 & 0.049 & 0.061 & 0.46 & 1.00 \\
        MSG & 3400 & 3000 & 30 & 0.36 & 5.0 & 21.1 & 7 & 0.000 & 0.013 & 0.46 & 1.00 \\
        MSS & 3400 & 3000 & 30 & 0.36 & 5.0 & 21.1 & 2 & 0.049 & 0.062 & 0.46 & 1.00 \\
        \hline
    \end{tabular}
\end{table*}

Time-dependent spot filling factors are \vone{initially estimated} using \texttt{spotter} \citep{garcia2025}. For each spectral type and spot size, a peak-to-peak variability amplitude of 1\% in the VISDA band is adopted \vone{for the \texttt{spotter} simulations}, consistent with typical rotational modulation in active K- and M-dwarfs. Circular spots are distributed isotropically over the stellar surface and the number of spots is increased until the VISDA rotational modulation reaches 1\%. These geometric assumptions are used only to provide representative spot-covering-fraction ranges and are not intended to model a true stellar surface. Our retrieval framework only uses the resulting disk-integrated spectra and projected spot covering fractions rather than the detailed spot geometry. For each configuration, minimum and maximum projected spot filling factors over a rotation cycle, $f_{\rm min}$ and $f_{\rm max}$, are calculated by averaging over $100$ realizations. 

\vone{However, \texttt{spotter} applies a spatially varying contrast profile to each spot, whereas the subsequent Pandora data simulations assume a uniform spot contrast. 
As filling factors extracted from \texttt{spotter} include pixels in the low-contrast wings of the spot profiles, applying a uniform contrast across these areas therefore produces substantially larger VISDA variability amplitudes than the original 1\% target. 
For this reason, we rescaled the filling-factor ranges in this study to produce approximately 1\% peak-to-peak VISDA variability across the sampled pre- and post-transit binned observations.}

These values are used to define a sinusoidal time series $f_{\rm spot}(t)$ that varies between $f_{\rm min}$ and $f_{\rm max}$ over the adopted rotation period, describing the time evolution of the spot filling factor. The sinusoidal model is a simplifying choice that provides a controlled sampling of different stellar activity states, and more complex spot-covering-fraction models can be implemented, as discussed in Section~\ref{subsec:future_extensions}.

The disk-integrated stellar spectrum at each time step $t$ is constructed as a linear combination of a quiet photosphere and a cool spot component both drawn from the PHOENIX ACES grid \citep{phoenix},
\begin{equation}\label{eq:SC_eq}
    F_\star(t,\lambda) = \left[1 - f_{\rm spot}(t)\right] F_{\rm phot}(\lambda) + f_{\rm spot}(t)\,F_{\rm spot}(\lambda).
\end{equation}
$F_{\rm phot}(\lambda)$ and $F_{\rm spot}(\lambda)$ are the stellar surface fluxes at the photospheric and spot temperatures, respectively. The stellar spectra are scaled to yield the flux density at Earth and propagated through an instrument response and noise model using the \texttt{pandora-sat} package \citep{hedges2024}.

Each of the eight scenarios has 10 simulated visits to mimic a realistic Pandora dataset. Each visit spans a 24-hour window centered on a hypothetical transit, with an assumed planetary orbital period of three days \citep{barclay2025a}. Within each 24-hour visit, out-of-transit data are grouped into two 11-hour windows: a pre-transit bin and a post-transit bin, excluding data within $\pm 1$\,hr of transit center. This procedure yields 20 out-of-transit epochs per stellar scenario across eight scenarios, for a combined set of 160 binned datasets. Each dataset thus consists of one VISDA broadband point and one NIRDA spectrum corresponding to an 11-hour out-of-transit interval, along with the underlying injected stellar parameters and spot filling factors. In this work, we take these binned VISDA and NIRDA datasets as the input to our extended retrieval framework.

\begin{figure*}
    \centering
    \includegraphics[width=1\linewidth]{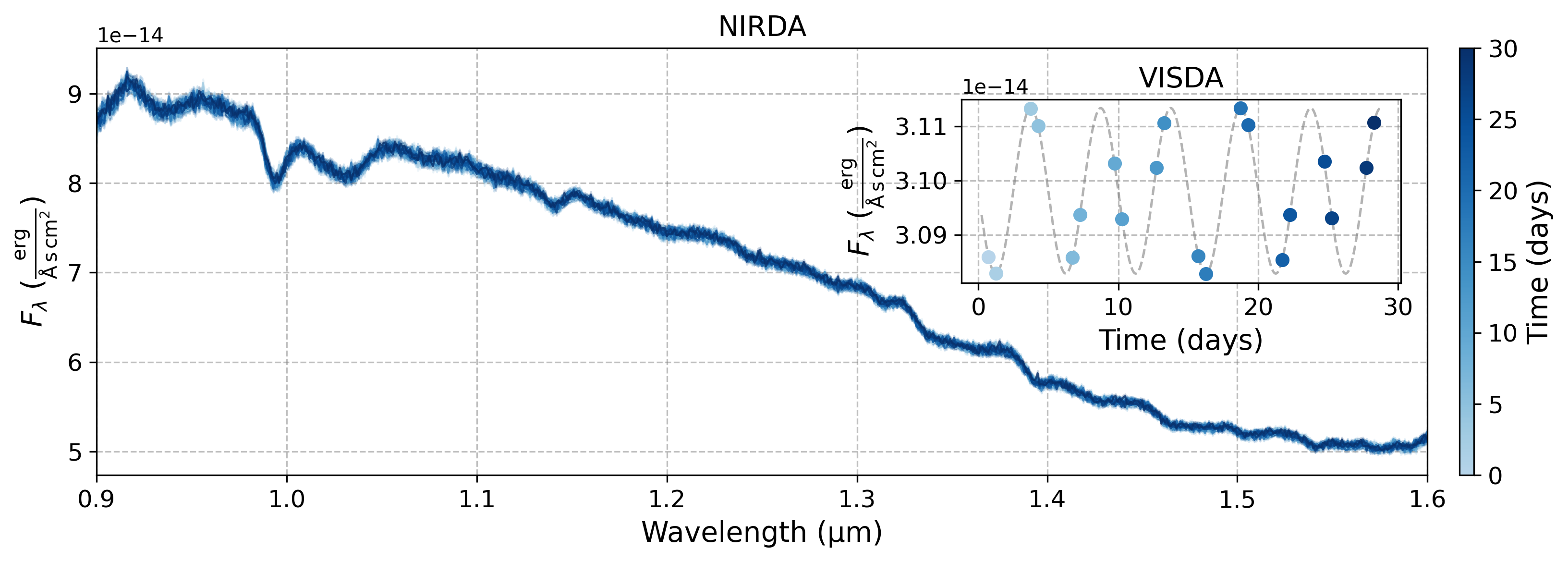}
    \caption{We simulate out-of-transit Pandora VISDA and NIRDA observations of representative targets as inputs to our retrieval analyses. Here, we show the M-dwarf, fast-rotator, giant-spot scenario. Each epoch consists of one VISDA broadband photometric point and one NIRDA spectrum, binned over an 11-hour pre- or post-transit interval. For each of the eight stellar scenarios, the simulated dataset contains 10 24-hour visits, corresponding to 20 out-of-transit epochs, with colors indicating time of observation.}
    \label{fig:data}
\end{figure*}

\section{Methods I: Stellar Retrievals} \label{sec:retrieval-set-up}

\subsection{Baseline Model} \label{subsec:baseline}

To simulate the effects of stellar model--data mismatch, we generate and retrieve the stellar spectra using different PHOENIX model grids. The simulated Pandora observations described in Section~\ref{sec:simulated} are generated with the PHOENIX ACES grid \citep{phoenix}, while the retrievals in this work use the PHOENIX NewEra grid \citep{newera}. \vone{As a control, we also retrieve the same simulated observations with the PHOENIX ACES grid used to generate them, so that no model--data mismatch is present; these retrievals are used only for the model comparison in Section~\ref{subsec:when_to_use}.} Both grids are 1D local thermodynamic equilibrium (LTE) stellar atmosphere models with similar parameter coverage, but they differ primarily in their opacity inputs and model updates \citep{newera}. Because molecular opacities dominate the spectra of cool K- and M-dwarfs, the largest ACES-NewEra discrepancies are expected at low effective temperatures. This setup provides a controlled proxy for the real observational case in which the molecular complexity of cool photospheres and magnetically active regions is not perfectly captured by the adopted stellar atmosphere model.

\begin{figure*}[t]
    \centering
    \includegraphics[width=1.0\linewidth]{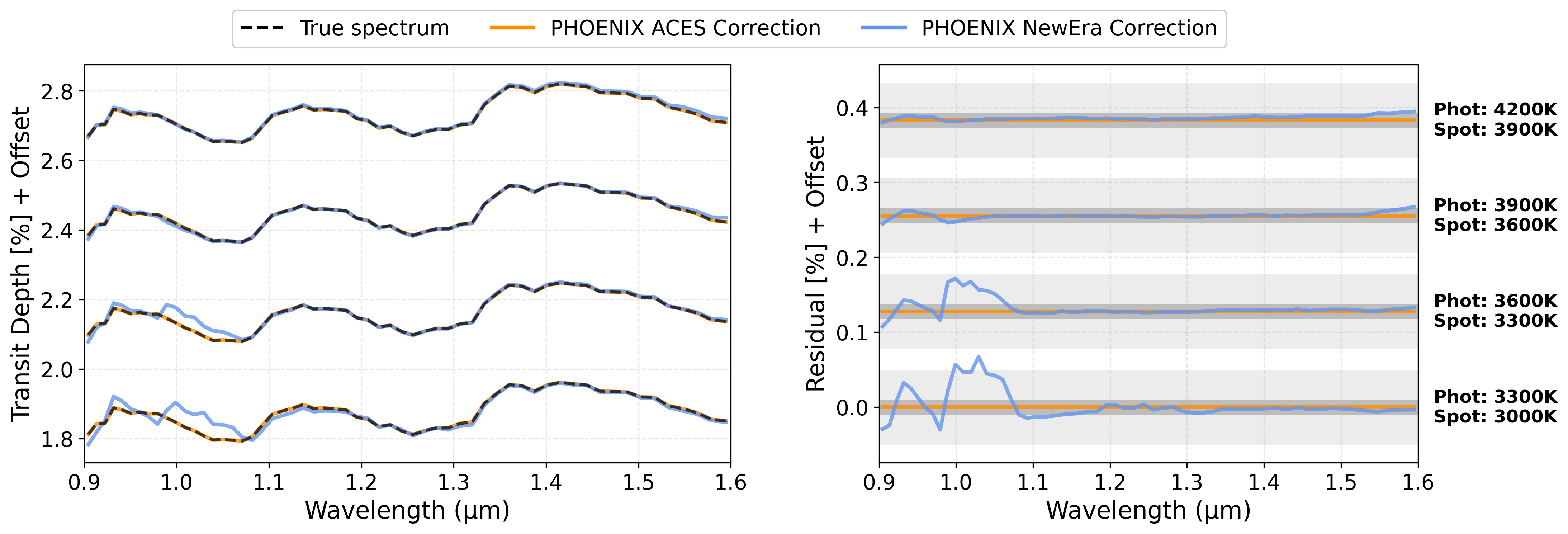}
    \caption{Effect of ACES-NewEra stellar model mismatch on contamination-correction accuracy. The left panel shows simulated water-bearing transmission spectra after correcting for stellar contamination using either the correct PHOENIX ACES grid or the mismatched PHOENIX NewEra grid. The true uncontaminated spectrum (black dashed) is shown alongside the corrected spectra for each temperature pair. The right panel shows the residuals between the corrected and true transmission spectra. The dark gray shaded regions indicate 100~ppm uncertainty, a conservative representation of JWST instrumental precision \citep{espinoza2026}. At $R=30$, Pandora is expected to reach precisions of $\sim100$~ppm for bright targets ($m_J \approx 7$), but for fainter targets ($m_J \gtrsim 10$), the uncertainty is expected to be $\sim500$~ppm \citep{rotman2026}, which is shaded in light gray.}
    \label{fig:stellar-model-comparison}
\end{figure*}

Figure~\ref{fig:stellar-model-comparison} demonstrates how this ACES-NewEra mismatch can propagate into a stellar-contamination correction. We generate an uncontaminated transmission spectrum representative of a Pandora target using \texttt{POSEIDON} \citep{macdonald2017, macdonald2023poseidon}, setting a water abundance of $\log X_{\rm H_2O} = -2$ over the spectral range and resolution of Pandora's NIRDA detector (0.9 to 1.6~$\micron$, $R \sim 120$). We inject a stellar contamination instance representative of a Pandora target using the PHOENIX ACES stellar library, following the standard two-component contamination prescription \citep{rackham2018a}. We then attempt to recover the uncontaminated transmission spectrum by correcting for stellar contamination using either the correct PHOENIX ACES grid or the mismatched PHOENIX NewEra grid at the same physical parameters. When the correction uses the correct ACES grid, the uncontaminated transmission spectrum is fully recovered; however, when the correction uses the mismatched NewEra grid, wavelength-dependent residuals persist, and the discrepancy grows toward cooler stellar temperatures. For temperatures below $\sim$3600\,K, the discrepancies can exceed 100~ppm, suggesting that this discrepancy can fundamentally limit the accuracy of contamination corrections. We therefore first define the deterministic stellar-contamination model used as the baseline for comparison.

The baseline stellar contamination model follows the two-component prescription given in Equation~\ref{eq:SC_eq}. The disk-integrated stellar spectrum is modeled as a linear combination of a quiescent photosphere spectrum and a cool spot spectrum, weighted by the spot covering fraction. \vone{For both the non-GP and GP-aided retrievals, we model the spot-covering fraction time series sinusoidally as}

\begin{equation}
    f_{\rm spot}(t)
    =
    f_{\rm mean}
    +
    A\sin\left(2\pi t/P_{\rm rot}+\phi\right),
    \label{eq:fspot_sinusoid}
\end{equation}

\vone{where $f_{\rm mean}$ is the mean covering fraction, $A$ is the modulation semi-amplitude, $P_{\rm rot}$ is the stellar rotation period, and $\phi$ is the phase.} We intentionally adopt this simplified prescription, \vone{matching the functional form used to generate the simulated dataset,} to isolate the wavelength-domain model--data mismatch problem without adding another layer of time-domain model flexibility. Future extensions are discussed in Section~\ref{subsec:future_extensions}.

\vone{This sinusoidal form highlights an identifiability limitation of the two-component mixture model. Substituting Equation~\ref{eq:fspot_sinusoid} into Equation~\ref{eq:SC_eq} shows that the time-dependent disk-integrated spectrum can be written as}

\begin{equation}
    F_\star(t,\lambda)
    =
    M(\lambda)
    +
    V(\lambda)
    \sin\left(2\pi t/P_{\rm rot}+\phi\right),
    \label{eq:mean_modulation_decomposition}
\end{equation}
where
\begin{align}
    M(\lambda)
    &=
    \left[1-f_{\rm mean}\right]F_{\rm phot}(\lambda)
    +
    f_{\rm mean}F_{\rm spot}(\lambda),
    \label{eq:mean_spectrum_decomposition} \\
    V(\lambda)
    &=
    A\left[F_{\rm spot}(\lambda)-F_{\rm phot}(\lambda)\right].
    \label{eq:modulation_spectrum_decomposition}
\end{align}

\vone{The observations therefore constrain the mean composite spectrum $M(\lambda)$ and the modulation spectrum $V(\lambda)$ directly, while the individual component spectra and spot covering fractions are inferred quantities.}

The model is parameterized by the photosphere temperature $T_{\rm phot}$, the temperature contrast $\Delta T = T_{\rm phot} - T_{\rm spot}$, the spot covering fraction $f_{\rm spot}(t)$, and the stellar radius $R_\star$. Priors for all physical model parameters are summarized in Table~\ref{tab:priors}. We enforce $T_{\rm spot} < T_{\rm phot}$ and adopt small, nonzero lower bounds on the mean spot covering fraction $f_{\rm mean}$ and its modulation amplitude $A$ to preserve the two-component, time-variable model. $f_{\rm mean}=0$ would correspond to no spotted component, and $A=0$ would correspond to no rotationally modulated  mixture. These bounds do not affect the simulated scenarios considered, as the smallest true values are \vone{$f_{\rm mean}=0.0055$ and $A=0.0055$.} We note that the instantaneous filling factor can still reach $f_{\rm spot}(t)=0$. 

\begin{table*}
    \centering
    \caption{Prior distributions for all retrieval parameters. The physical-parameter priors define the deterministic stellar model used to construct the GP mean functions. The resulting physical parameters are not interpreted as final stellar constraints because they may be biased by stellar model--data mismatch. The same physical-parameter priors are used for the non-GP and GP-aided retrievals. The GP hyperparameter priors apply independently to both the photosphere and spot GPs.}
    \label{tab:priors}
    \begin{tabular}{lcc}
    \hline
    \hline
    Parameter & Prior & Bounds \\
    \hline
    \multicolumn{3}{c}{\textit{Physical Parameters}} \\
    $T_{\rm phot}$ (K), K-dwarf scenarios 
        & $\mathcal{U}$ 
        & $3000,5000$ \\
    $T_{\rm phot}$ (K), M-dwarf scenarios 
        & $\mathcal{U}$ 
        & $2300,4300$ \\
    $\Delta T = T_{\rm phot}-T_{\rm spot}$ (K), K-dwarf scenarios 
        & $\mathcal{U}$ 
        & $0,T_{\rm phot}-3000$ \\
    $\Delta T = T_{\rm phot}-T_{\rm spot}$ (K), M-dwarf scenarios 
        & $\mathcal{U}$ 
        & $0,T_{\rm phot}-2300$ \\
    $f_{\rm mean}$ 
        & $\mathcal{U}$ 
        & $0.001,0.85$ \\
    $A$ 
        & $\mathcal{U}$ 
        & $0.001,\min(0.35, f_{\rm mean}, 1-f_{\rm mean})$ \\
    $\phi$ (rad) 
        & $\mathcal{U}$ 
        & $0,2\pi$ \\
    $P_{\rm rot}$ (d) 
        & $\mathcal{U}$ 
        & $1,50$ \\
    $R_\star$ ($R_\odot$), K-dwarf scenarios 
        & $\mathcal{N}_{\rm trunc}(0.73,0.03\times0.73)$ 
        & $0.6,0.9$ \\
    $R_\star$ ($R_\odot$), M-dwarf scenarios 
        & $\mathcal{N}_{\rm trunc}(0.36,0.03\times0.36)$ 
        & $0.25,0.5$ \\
    \hline
    \multicolumn{3}{c}{\textit{GP Hyperparameters}} \\
    $\ln \sigma^2$ ($\ln$(erg s$^{-1}$ cm$^{-2}$ \AA$^{-1}$)$^2$)
        & $\mathcal{U}$ 
        & $-80,-65$ \\
    $\ln \ell$ ($\ln$($\micron$))
        & $\mathcal{U}$ 
        & $-7,0$ \\
    \hline
\end{tabular}
    
    \vspace{0.5em}
    \begin{flushleft}
    \footnotesize
    Notes. $\mathcal{U}$ denotes a uniform prior and $\mathcal{N}_{\rm trunc}(\mu,\sigma)$ denotes a normal prior with mean $\mu$ and standard deviation $\sigma$, truncated to the listed bounds. 
    The spot covering fraction is parameterized as $f_{\rm spot}(t)=f_{\rm mean}+A\sin(2\pi t/P_{\rm rot}+\phi)$, and the amplitude prior upper bound is set such that $f_{\rm spot}(t)$ remains within $(0,1)$.
    \end{flushleft}
\end{table*}

\vone{The model is evaluated on the combined VISDA+NIRDA data: the 701 NIRDA spectral bins and a single VISDA broadband point. For NIRDA, the component surface fluxes are passed through the NIRDA instrument response and binned to the detector wavelength grid with pandora-sat \citep{hedges2024}. For VISDA, the surface flux of each component is integrated over the full VISDA sensitivity curve (0.4–0.7 $\mu$m) and converted to a band-averaged flux density, also with pandora-sat, consistent with how the simulated observations were generated.}

To verify that our simulated model-model mismatch is representative of real model--data mismatch, we compare against empirical observations of TRAPPIST-1, a cool (2566\,K) 0.09~$M_\odot$ M-dwarf \citep{piaulet2025}. From two JWST/NIRISS/SOSS observations of TRAPPIST-1 b, the out-of-transit data do not match PHOENIX ACES and SPHINX models evaluated at literature values \citep{lim2023}. Both PHOENIX ACES and SPHINX are 1D LTE stellar atmosphere grids, though SPHINX is specifically designed for cooler M-dwarfs (2000--4000\,K) and incorporates updated molecular opacities relevant at these temperatures \citep{iyer2023}. The out-of-transit stellar spectral mismatch is approximately 10--20\% RMS across the NIRISS/SOSS wavelength range of 0.6--2.8~$\mu$m, with the optical (0.6--1.1~$\mu$m) dominating at $\sim$20--25\%. This stellar model infidelity propagates into the transmission spectrum, resulting in an estimated model-driven uncertainty of $\sim$1800~ppm at native pixel resolution \citep{lim2023}. Similarly, two NIRISS/SOSS observations of TRAPPIST-1 c show that the overall quality of out-of-transit observations fit with PHOENIX ACES models are poor, with $\chi_\nu^2 > 350$, and are especially poor under 1.1~$\mu$m \citep{radica2025}. IRTF/SpeX comparisons for mid-to-late M-dwarfs including TRAPPIST-1 also show structured model--data residuals at the 10--20\% level, with the largest discrepancies occurring \vone{at shorter wavelengths}, especially below 0.75~$\mu$m \citep{iyer2026}. These \vone{shorter-wavelength} regions are particularly sensitive to molecular opacity from FeH and TiO/VO, cloud opacity, convective treatment, and photospheric heterogeneity, all of which remain challenging to capture in 1D stellar atmosphere models.

Our ACES versus NewEra model-model mismatch at 3000\,K shows a similar pattern of 5--8\% RMS over the full range, with the optical dominating at 10--15\%. The TRAPPIST-1 spectra show the largest discrepancies with the models in the 1.0--1.1~$\mu$m region \citep{lim2023, Davoudi2024}, which we also observe in our model comparisons (Figure~\ref{fig:stellar-model-comparison}). \vone{Similar discrepancies have been observed for the K-dwarf HAT-P-11, where PHOENIX NewEra models fit near-infrared spectra but struggle to reproduce optical observations \citep{Niraula2026}. These observations motivate our controlled test of how GP-aided retrievals can mitigate stellar-model inaccuracies, although model--data discrepancies may differ in magnitude and structure.}

\subsection{GP-aided Model} \label{subsec:gp-framework}

To account for the wavelength-dependent mismatch between the stellar models and the simulated observations, we introduce two GP components into the stellar contamination model: one associated with the photosphere spectrum and one associated with the spot spectrum. \vone{The GP-aided model uses the same two-component stellar contamination model and sinusoidal spot covering model described in Section~\ref{subsec:baseline} but allows the component spectra to deviate from the adopted stellar grid. The changing spot covering fraction across epochs provides leverage to constrain these latent component spectra, but the identifiability limitation described in Section~\ref{subsec:baseline} remains: the data constrain the composite spectra more directly than the individual photosphere spectrum, spot spectrum, and spot covering fractions.} 

\vone{The deterministic non-GP retrieval strictly follows the stellar grids as exact models, which strongly constrains the photosphere and spot components but can bias the inferred spectra and filling factors when the grids are inaccurate. The GP-aided retrieval instead treats the stellar grids as a physically motivated mean function and allows wavelength-correlated deviations from it. This added flexibility reduces model-driven bias when the stellar grid is inaccurate, but relaxing the grid constraint also makes the latent components more dependent on the chosen models and priors. The GP-corrected spectra should therefore be interpreted as data-constrained latent components rather than fully model-independent measurements of the true stellar components.}

Since the data are fit with the incorrect deterministic stellar model, the resulting baseline physical parameters are expected to be biased and are not interpreted as stellar constraints. We write our model as a joint Gaussian distribution over the component spectra and the observed data, and we use the corresponding conditional distribution to infer the GP-corrected photosphere and spot spectra.

For each out-of-transit epoch $i$, we write the spectrum observed by Pandora as
\begin{equation}
    \mathbf{y}_i =
    (1 - f_i)\,\mathbf{p}
    + f_i\,\mathbf{s}
    + \mathbf{n}_i,
    \label{eq:epoch_model_gp}
\end{equation}
where $\mathbf{y}_i$ is the observed VISDA+NIRDA data at epoch $i$, $f_i$ is the spot covering fraction at epoch $i$, $\mathbf{p}$ is the GP-corrected photosphere spectrum, $\mathbf{s}$ is the GP-corrected spot spectrum, and $\mathbf{n}_i$ is the measurement noise at epoch $i$. Each spectrum is evaluated on the combined VISDA+NIRDA wavelength grid. \vone{Each VISDA flux value is band-averaged, obtained by integrating the model spectrum over the VISDA throughput. The GP VISDA correction is therefore also averaged over the VISDA throughput, such that each GP contributes a single additive correction to the band-averaged VISDA flux. This correction is shared across epochs and correlated with the NIRDA corrections through the kernel.}

The component spectra are modeled as Gaussian processes about deterministic stellar mean spectra,
\begin{align}
    \mathbf{p} &\sim
    \mathcal{N}\left(\boldsymbol{\mu}_{\rm phot}, \mathbf{K}_{\rm phot}\right), \\
    \mathbf{s} &\sim
    \mathcal{N}\left(\boldsymbol{\mu}_{\rm spot}, \mathbf{K}_{\rm spot}\right),
    \label{eq:component_gp_priors}
\end{align}
where $\boldsymbol{\mu}_{\rm phot}$ and $\boldsymbol{\mu}_{\rm spot}$ are the deterministic photosphere and spot mean spectra, and $\mathbf{K}_{\rm phot}$ and $\mathbf{K}_{\rm spot}$ are the corresponding GP covariance matrices. \vone{We assume the photosphere and spot GP deviations are independent and discuss this simplifying assumption in Section~\ref{subsec:future_extensions}.} The GP is intended to represent imperfectly modeled opacity structure, which may manifest as narrow molecular-opacity structures rather than smooth continuum offsets. Therefore, we adopt a Mat\'{e}rn-1/2, or exponential, covariance kernel over the combined VISDA+NIRDA wavelength grid for both components. This kernel produces continuous but non-differentiable functions, making it less smooth than commonly used squared-exponential or higher order Mat\'ern kernels. The GP-aided model therefore introduces four spectral GP hyperparameters: the photosphere GP variance amplitude $\sigma^2_{\rm phot}$ and correlation length scale $\ell_{\rm phot}$, and the spot GP variance amplitude $\sigma^2_{\rm spot}$ and correlation length scale $\ell_{\rm spot}$. The length-scale priors span from the spacing between adjacent wavelength points to the full combined VISDA+NIRDA wavelength range. \vone{The amplitude priors are informed by the residuals of the deterministic non-GP retrievals across all scenarios: the range is chosen to broadly cover the residual amplitudes of every scenario, so the same priors are adopted across scenarios.} The mean functions $\boldsymbol{\mu}_{\rm phot}$ and $\boldsymbol{\mu}_{\rm spot}$ depend on the parameters of the deterministic photosphere and spot models described in Section~\ref{subsec:baseline}, which are inferred simultaneously with the GP hyperparameters. All priors are summarized in Table~\ref{tab:priors}.

The deterministic mean model for epoch $i$ is
\begin{equation}
    \mathbf{m}_i =
    (1 - f_i)\,\boldsymbol{\mu}_{\rm phot}
    + f_i\,\boldsymbol{\mu}_{\rm spot}.
    \label{eq:epoch_mean}
\end{equation}

Assuming independence between the photosphere GP, the spot GP, and the measurement noise, the covariance between the spectra observed at epochs $i$ and $j$ is
\begin{equation}
    \mathbf{C}_{ij}
    =
    (1 - f_i)(1 - f_j)\,\mathbf{K}_{\rm phot}
    + f_i f_j\,\mathbf{K}_{\rm spot}
    + \delta_{ij}\boldsymbol{\Sigma}_i,
    \label{eq:cov_block}
\end{equation}
where $\boldsymbol{\Sigma}_i$ is the measurement-noise covariance matrix for epoch $i$. The first term describes the covariance from the shared photosphere GP, the second term describes covariance from the shared spot GP, and the final term accounts for independent measurement noise.

The GP-aided stellar model can thus be written as the joint Gaussian distribution over the photosphere spectrum $\mathbf{p}$, spot spectrum $\mathbf{s}$, and observed spectra $\mathbf{y}$:
\begin{equation}
    \begin{bmatrix}
    \mathbf{p} \\
    \mathbf{s} \\
    \mathbf{y}
    \end{bmatrix}
    \sim
    \mathcal{N}
    \left(
    \begin{bmatrix}
    \boldsymbol{\mu}_{\rm phot} \\
    \boldsymbol{\mu}_{\rm spot} \\
    \mathbf{m}
    \end{bmatrix},
    \begin{bmatrix}
    \mathbf{K}_{\rm phot} & 0 & \mathbf{K}_{{\rm phot},y} \\
    0 & \mathbf{K}_{\rm spot} & \mathbf{K}_{{\rm spot},y} \\
    \mathbf{K}_{{\rm phot},y}^{\rm T} & \mathbf{K}_{{\rm spot},y}^{\rm T} & \mathbf{C}
    \end{bmatrix}
    \right)
    \label{eq:joint_gp_distribution}
\end{equation}
where $\mathbf{y}$ is the stacked observed Pandora data from Equation~\ref{eq:epoch_model_gp}, $\mathbf{m}$ is the corresponding stacked deterministic mean model from Equation~\ref{eq:epoch_mean}, and $\mathbf{C}$ is the full data covariance matrix from Equation~\ref{eq:cov_block}. The cross-covariance matrices between the component spectra and the stacked data are
\begin{align}
    \mathbf{K}_{{\rm phot},y}
    &=
    \begin{bmatrix}
    (1 - f_0)\mathbf{K}_{\rm phot} &
    \cdots &
    (1 - f_{n-1})\mathbf{K}_{\rm phot}
    \end{bmatrix}, \\
    \mathbf{K}_{{\rm spot},y}
    &=
    \begin{bmatrix}
    f_0\mathbf{K}_{\rm spot} &
    \cdots &
    f_{n-1}\mathbf{K}_{\rm spot}
    \end{bmatrix}.
    \label{eq:cross_covariances}
\end{align}
Equation~\ref{eq:joint_gp_distribution} defines the full GP-aided stellar model used in our retrievals.

Applying the standard conditional Gaussian identity gives the posterior distribution of the component spectra conditioned on the Pandora data,
\begin{equation}
    \mathbf{x}\mid\mathbf{y}
    \sim
    \mathcal{N}
    \left(
    \boldsymbol{\mu}_x
    +
    \mathbf{K}_{xy}\mathbf{C}^{-1}\mathbf{r},
    \,
    \mathbf{K}_{xx}
    -
    \mathbf{K}_{xy}\mathbf{C}^{-1}\mathbf{K}_{yx}
    \right),
    \label{eq:conditional_ps}
\end{equation}
where $\mathbf{x}=[\mathbf{p},\mathbf{s}]^{\rm T}$,
$\boldsymbol{\mu}_x=[\boldsymbol{\mu}_{\rm phot},
\boldsymbol{\mu}_{\rm spot}]^{\rm T}$, $\mathbf{r}=\mathbf{y}-\mathbf{m}$, and
\begin{align}
    \mathbf{K}_{xx}
    &=
    \begin{bmatrix}
    \mathbf{K}_{\rm phot} & 0 \\
    0 & \mathbf{K}_{\rm spot}
    \end{bmatrix}, \\
    \mathbf{K}_{xy}
    &=
    \begin{bmatrix}
    \mathbf{K}_{{\rm phot},y} \\
    \mathbf{K}_{{\rm spot},y}
    \end{bmatrix}, \\
    \mathbf{K}_{yx}
    &=
    \mathbf{K}_{xy}^{\rm T}.
    \label{eq:Kxx_Kxy}
\end{align}

We sample from Equation~\ref{eq:conditional_ps} to obtain posterior samples of the GP-corrected photosphere and spot spectra. For computational efficiency, the marginalized GP likelihood is evaluated using the Woodbury identity and the matrix determinant lemma, avoiding direct inversion of the full stacked data covariance matrix \citep{rasmussen2006}.

We run the non-GP retrieval and the GP-aided retrieval on all eight scenarios in the simulated Pandora dataset described in Section~\ref{sec:simulated}, using the full set of 20 out-of-transit epochs for each scenario. \vone{Stellar retrievals are performed with the nested sampling package UltraNest \citep{buchner2021} using its \texttt{ReactiveNestedSampler} with a slice step sampler ($10\,n_{\rm dim}$ steps per iteration).}

\section{Methods II: Atmospheric Retrievals} \label{sec:transmission-propagation}

To assess how the inferred stellar spectra propagate into atmospheric inferences, we generate a simulated planetary transmission spectrum, inject stellar contamination corresponding to each simulated Pandora stellar scenario, and compare atmospheric retrievals that marginalize over the contamination-factor posteriors inferred from the non-GP and GP-aided stellar retrievals.

\subsection{Simulating Transmission Spectra}
\label{subsec:true-transmission-spectrum}

Using \texttt{POSEIDON} \citep{macdonald2017, macdonald2023poseidon}, we model the simulated transmission spectrum after the gas giant WASP-69 b, a representative Pandora target \citep{Gilbert2026} with water features and stellar contamination effects \citep{khalafinejad2021, petit2024}. Table~\ref{tab:transmission-spectrum-params} summarizes the parameters used. The atmosphere is modeled as an H$_2$-He-dominated atmosphere with H$_2$O as the only trace absorber. The pressure-temperature profile is assumed to be isothermal, spanning $10^{-7}$ to $100$~bar, and the atmosphere is taken to be cloud-free. We compute the transmission spectrum over the Pandora/NIRDA wavelength range (0.9--1.6~$\micron$) at a resolving power of $R = 10{,}000$. \vone{We then bin this spectrum onto 58 logarithmically spaced bins at R=100 spanning 0.9 to $1.6~\mu$m, comparable to NIRDA's resolving power.} This simplified atmospheric setup is chosen to produce a controlled water-bearing transmission spectrum rather than to provide a complete atmospheric model of WASP-69 b.

\begin{table}
    \centering
    \caption{Parameters used to generate the uncontaminated WASP-69 b transmission spectrum with \texttt{POSEIDON}.}
    \label{tab:transmission-spectrum-params}
    \begin{tabular}{lc}
    \hline
    \hline
    Parameter & Value \\
    \hline
    \multicolumn{2}{c}{\textit{Stellar Parameters}} \\
    Stellar radius, $R_\star$ & $0.801\,R_\odot$$^{a}$ \\
    Stellar effective temperature, $T_{\rm eff}$ & $4792$ K$^{a}$ \\
    Stellar surface gravity, $\log g_\star$ (cgs) & $4.57$$^{a}$ \\
    Stellar metallicity, $[\mathrm{Fe/H}]$ & $0.35$$^{a}$ \\
    Apparent magnitude, $m_J$ & $9^{c}$ \\
    \hline
    \multicolumn{2}{c}{\textit{Planetary Parameters}} \\
    Planetary radius, $R_p$ & $1.0\,R_J$$^{a}$ \\
    Planetary mass, $M_p$ & $0.26\,M_J$$^{a}$ \\
    Equilibrium temperature, $T_{\rm eq}$ & $971$ K$^{a}$ \\
    \hline
    \multicolumn{2}{c}{\textit{Atmospheric Parameters}} \\
    H$_2$O volume mixing ratio, $\log X_{\rm H_2O}$ & $-2$$^{b}$ \\
    Pressure range & $10^{-7}$--$100$ bar \\
    Number of atmospheric layers, $N_{\rm layers}$ & $100$ \\
    Reference pressure, $P_{\rm ref}$ & $10$ bar \\
    Radius at reference pressure, $R_p(P_{\rm ref})$ & $1.0\,R_J$ \\
    \hline
    \multicolumn{2}{l}{$^{a}$\citet{allart2025}} \\
    \multicolumn{2}{l}{$^{b}$\citet{khalafinejad2021}} \\
    \multicolumn{2}{l}{$^{c}$\citet{rackham2026pand}} \\
    \end{tabular}
\end{table}

\subsection{Injecting Stellar Contamination}
\label{subsec:injection}

We inject stellar contamination into the uncontaminated planetary spectrum. Following \citet{rackham2018a}, the contamination factor is defined as
\begin{equation}
    \varepsilon(\lambda) = \frac{F_{\rm phot}(\lambda)}{(1 - f_{\rm spot}) F_{\rm phot}(\lambda) + f_{\rm spot} F_{\rm spot}(\lambda)},
    \label{eq:contamination-factor}
\end{equation}
where $F_{\rm phot}(\lambda)$ and $F_{\rm spot}(\lambda)$ are the photosphere and spot surface flux spectra, and $f_{\rm spot}$ is the spot covering fraction at the time of transit. \vone{For the non-GP retrieval, $F_{\rm phot}$ and $F_{\rm spot}$ are the PHOENIX NewEra photosphere and spot spectra evaluated at each posterior sample. For the GP-aided retrieval, they are the GP-corrected component spectra $\mathbf{p}$ and $\mathbf{s}$ from Equation~\ref{eq:epoch_model_gp}. In both cases, $f_{\rm spot}$ is the posterior spot-covering-fraction model evaluated at the transit epoch.}

This $\varepsilon$ factor represents the multiplicative bias imposed on transmission spectra by unocculted stellar surface heterogeneities. We consider only unocculted spots and assume that any spot-crossing events are either absent or have been identified and masked before constructing the transmission spectra. Because $f_{\rm spot}$ varies with time as the star rotates, each transit has a different contamination factor.

We compute the true contamination factor $\varepsilon_{\rm true}(\lambda)$ using the PHOENIX ACES photosphere and spot spectra and the true spot covering fraction at each transit epoch from the simulated dataset. The contaminated transmission spectrum is then
\begin{equation}
    D_{\rm contaminated}(\lambda) = D_{\rm true}(\lambda) \times \varepsilon_{\rm true}(\lambda) + n(\lambda),
    \label{eq:contaminated-spectrum}
\end{equation}
where $D_{\rm true}(\lambda)$ is the uncontaminated planetary spectrum and $n(\lambda)$ is \vone{a Gaussian noise realization drawn from the expected single-transit NIRDA precision, simulated with pandora-sat \citep{hedges2024} for a WASP-69 b-like system at m$_J$ = 9. The per-bin uncertainties range from 520 to 960 ppm (median 630 ppm) and are assumed uncorrelated between bins. We draw an independent noise realization for each of the 10 transits in each scenario. The non-GP and GP-aided atmospheric retrievals are applied to the same noisy spectra so that differences between them are not caused by different noise draws.}

Although the injected planetary spectrum is based on WASP-69 b, which has a $\sim4800$\,K K-dwarf host star, we apply stellar contamination factors from each of the eight simulated scenarios described in Section~\ref{sec:simulated}. This therefore does not model WASP-69 b's specific host star in each case, but rather uses a representative water-bearing gas-giant transmission spectrum to test how different stellar activity scenarios affect the contamination correction and atmospheric retrieval. We apply this procedure to each of the 10 transits in each of the eight stellar scenarios, producing contaminated transmission spectra that span the range of stellar activity levels expected for Pandora targets.

\subsection{Atmospheric Retrievals}
\label{subsec:atmospheric-retrievals}

To determine how stellar-model treatment propagates into atmospheric inference, we perform two atmospheric retrievals for each 10-transit scenario. In one retrieval, we use the contamination-factor posterior from the non-GP stellar retrieval; in the other, we use the contamination-factor posterior from the GP-aided stellar retrieval. \vone{We treat the stellar contamination factor as a nuisance quantity constrained by the stellar retrieval and marginalize over it in the atmospheric retrieval likelihood.} We use a simplified atmospheric setup to isolate the effect of our stellar-contamination corrections on inferred water abundance rather than to provide a complete atmospheric analysis.

From each stellar retrieval, \vone{we draw 1000 equally-weighted posterior samples produced by UltraNest, resampled in proportion to their nested-sampling posterior weights. Each draw gives a photosphere spectrum, a spot spectrum, and a spot-covering-fraction time series, from which we compute $\varepsilon(\lambda)$ from Equation~\ref{eq:contamination-factor} at each of the 10 transits. Each transit spectrum contains 58 wavelength bins, so each contamination-factor draw has $10\times58=580$ elements. This representation preserves the wavelength-to-wavelength and epoch-to-epoch correlations in the contamination factor learned by the stellar retrieval.}

\vone{Directly sampling all 580 elements of the contamination factor as independent nuisance parameters would be impractical and would discard the encoded correlation structure. We therefore represent the contamination posterior in a reduced basis. We compute the empirical mean contamination vector, $\boldsymbol{\mu}_\varepsilon$, from the 1000 stored draws and perform a singular value decomposition of the centered contamination-factor draws. We then write a contamination realization as}

\begin{equation}
    \boldsymbol{\varepsilon}(\mathbf{a})
    =
    \boldsymbol{\mu}_\varepsilon
    +
    \sum_{k=1}^{K} a_k \mathbf{v}_k,
    \label{eq:epsilon_pca}
\end{equation}

\vone{where $\mathbf{v}_k$ are the principal-component vectors and $a_k$ are the corresponding amplitudes sampled jointly with the atmospheric parameters. The prior on each coefficient $a_k$ is defined empirically from the stellar posterior. We project the stored contamination-factor draws onto each principal component and use the resulting empirical distribution as the prior on $a_k$.}

\vone{We choose the number of components using a signal-to-noise criterion. For each principal component, we estimate the size of the change in the \vtwo{modeled} transmission spectrum produced by a $1\sigma$ shift along that component relative to the observational uncertainties and retain components with ${\rm S/N}_k>1$. After projecting the stored contamination-factor draws onto the retained components, we compute the covariance of the residual contamination draws, \vtwo{$\boldsymbol{\Sigma}_\varepsilon^{\rm res}$. This describes the remaining contamination uncertainty, including its correlations across wavelength and across transits.}}

\vtwo{We stack the 10 contaminated transmission spectra into a single vector, $\mathbf{D}_{\rm obs}$, and define the stellar-contaminated model as $\mathbf{D}(\boldsymbol{\theta})\odot\boldsymbol{\varepsilon}(\mathbf{a})$, where $\odot$ denotes element-wise multiplication. We write the log-likelihood as}
\begin{equation}
\begin{aligned}
  \ln\mathcal{L}
  &=
  -\frac{1}{2}\mathbf{R}^{\rm T}\mathbf{C}^{-1}\mathbf{R}
  - \frac{1}{2}\ln\left|2\pi\mathbf{C}\right|, \\
  \mathbf{R}
  &=
  \mathbf{D}_{\rm obs}
  - \mathbf{D}(\boldsymbol{\theta})
    \odot \boldsymbol{\varepsilon}(\mathbf{a}), \\
  \mathbf{C}
  &=
  {\rm diag}(\boldsymbol{\sigma}^2)
  + {\rm diag}[\mathbf{D}(\boldsymbol{\theta})]\,
    \boldsymbol{\Sigma}_\varepsilon^{\rm res}\,
    {\rm diag}[\mathbf{D}(\boldsymbol{\theta})].
\end{aligned}
\label{eq:atm_like}
\end{equation}

\vtwo{$\mathbf{R}$ is the difference between the observed spectra and contaminated model spectra, and $\boldsymbol{\sigma}$ are the observational uncertainties. The second term in $\mathbf{C}$ propagates the residual contamination into transit-depth space, since $\boldsymbol{\varepsilon}$ affects the transit model multiplicatively.}

\begin{figure}
    \centering
    \includegraphics[width=\linewidth]{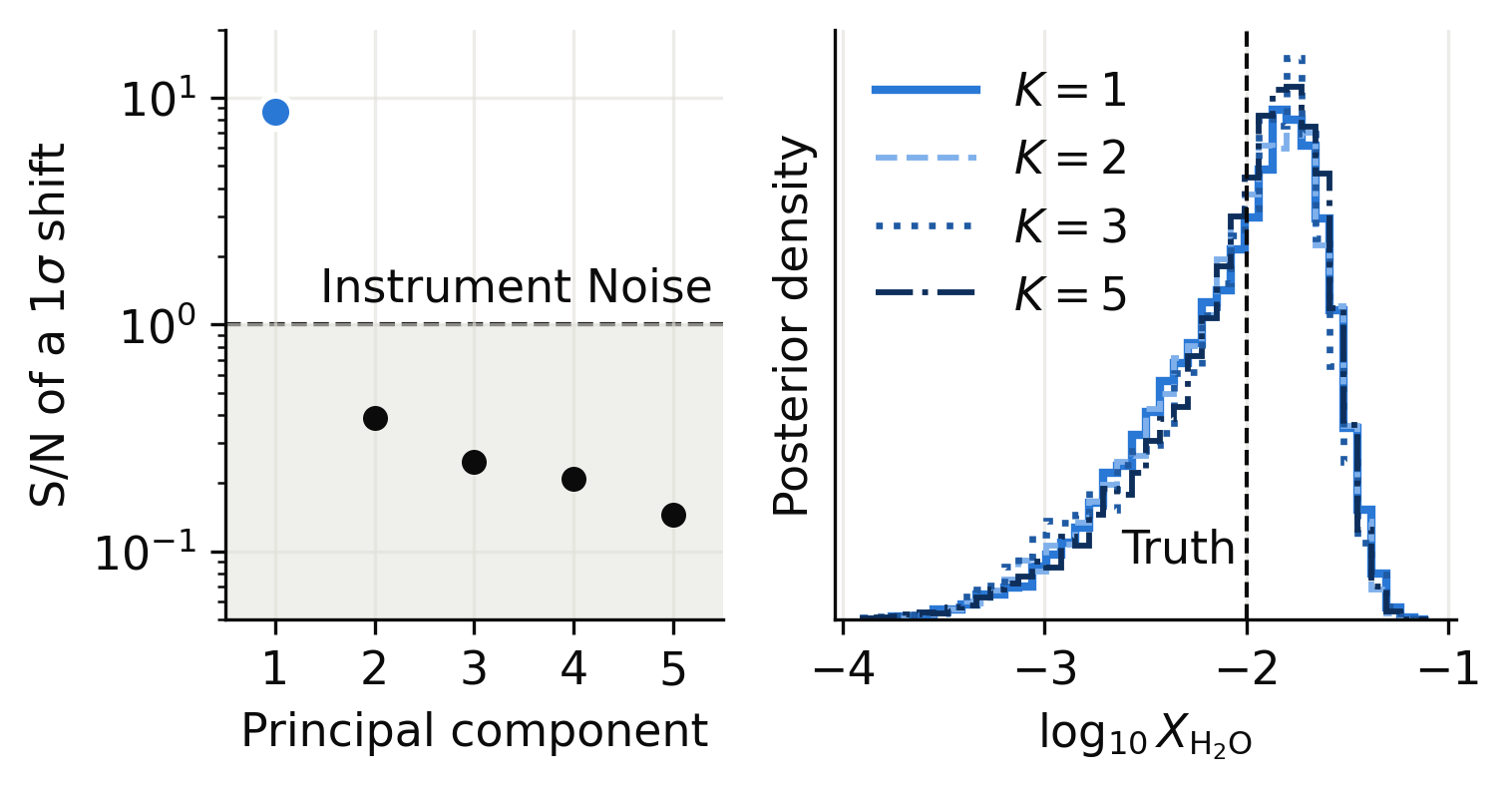}
    \caption{\vone{Test of the number of retained principal components of the stellar contamination factor for the M-dwarf fast-rotator giant-spot scenario. Left: signal-to-noise values for the first five contamination principal components. Only the leading component exceeds the adopted ${\rm S/N}_k>1$ threshold. Right: atmospheric retrieval posteriors obtained when retaining $K=1$, 2, 3, and 5 principal components. The posteriors are insensitive to the number of retained components, further validating our choice of $K=1$.}}
    \label{fig:pca_k_test}
\end{figure}

For all scenarios, only the leading principal component satisfies the SNR criterion, so we adopt $K=1$ in all atmospheric retrievals. We further validate this by running atmospheric retrievals with $K=1$, 2, 3, and 5 for a representative scenario and find consistent retrieved atmospheric posteriors. These tests are shown in Figure~\ref{fig:pca_k_test}.

\vone{Through this reduced-basis representation, the atmospheric retrieval samples both the atmospheric parameters and the stellar-contamination nuisance parameter. The reported atmospheric posteriors are marginalized over the contamination parameter, and therefore incorporate the stellar-contamination uncertainty propagated from the stellar retrieval.}

We use \texttt{POSEIDON} \citep{macdonald2017, macdonald2023poseidon} with \texttt{dynesty}'s Dynamic Nested Sampler \citep{speagle2020dynesty}, using 500 live points. The priors used in the atmospheric retrievals are summarized in Table~\ref{tab:atm-retrieval-priors}. We use the same atmospheric retrieval setup with both the non-GP and GP-aided contamination factor posteriors. 

\begin{table}
\centering
\caption{Atmospheric retrieval parameters and prior bounds.}
\label{tab:atm-retrieval-priors}
\footnotesize
\setlength{\tabcolsep}{3pt}
\begin{tabular}{lccc}
\hline
\hline
Parameter & Lower & Upper & Prior \\
          & Bound & Bound & Type \\
\hline
\parbox[t]{1.4in}{\raggedright Atmospheric temperature $T$ [K]}
& 500 & 1500 & Uniform \\
\parbox[t]{1.4in}{\raggedright Log H$_2$O volume mixing ratio $\log X_{\rm H_2O}$}
& $-12$ & $-1$ & Uniform \\
\parbox[t]{1.4in}{\raggedright Log reference pressure $\log P_{\rm ref}$ [bar]}
& $-6$ & $2$ & Uniform \\
\hline
\end{tabular}
\end{table}

We assume an H$_2$-He-dominated atmosphere with H$_2$O \citep{polyansky2018exomol} as the only trace absorber. The atmosphere is cloud-free and haze-free, with an isothermal pressure-temperature profile. The He abundance is fixed to its solar value relative to H$_2$, such that $\mathrm{He}/\mathrm{H}_2 = 0.17$. The choice to retrieve the reference radius or pressure does not meaningfully impact the retrieved precision on the molecular abundances \citep{welbanks2019}. Pandora targets are transiting planets with precise radius measurements from prior observations, so we fix the planetary reference radius to the input planetary radius listed in Table~\ref{tab:transmission-spectrum-params} and retrieve the reference pressure $P_{\rm ref}$. The remaining stellar and planetary parameters are also fixed to the values used to generate the input spectrum, listed in Table~\ref{tab:transmission-spectrum-params}. The free atmospheric parameters are thus the temperature $T$, the H$_2$O volume mixing ratio $X_{\rm H_2O}$, and the reference pressure $P_{\rm ref}$. Model spectra are computed from 0.7--1.8~$\micron$ at a constant resolving power of $R=10{,}000$ and binned to the wavelength grid of the Pandora NIRDA transmission spectrum.

\section{Results}\label{sec:results}
We fit the simulated out-of-transit stellar spectra for each scenario using both the deterministic non-GP stellar retrieval described in Section~\ref{subsec:baseline} and the GP-aided stellar retrieval described in Section~\ref{subsec:gp-framework}. In Section~\ref{subsec:results-example-msg}, we present detailed results for one representative scenario. In Section~\ref{subsec:results-transmission-correction}, we present the contamination-corrected transmission spectra and atmospheric retrieval results for the same example scenario. Finally, in Section~\ref{subsec:results-all-scenarios}, we summarize the retrieval performance across all eight simulated Pandora scenarios.

\subsection{Example Scenario: M-dwarf fast rotator with giant spots}
\label{subsec:results-example-msg}

We first present detailed results for the M-dwarf fast-rotator giant-spot (MFG) scenario. The cool M-dwarf temperatures produce the largest spectral mismatch between the PHOENIX ACES and PHOENIX NewEra grids, while the changing spot covering fraction across epochs helps separate the photosphere and spot contributions.

\begin{figure*}
    \centering
    \includegraphics[width=1\linewidth]{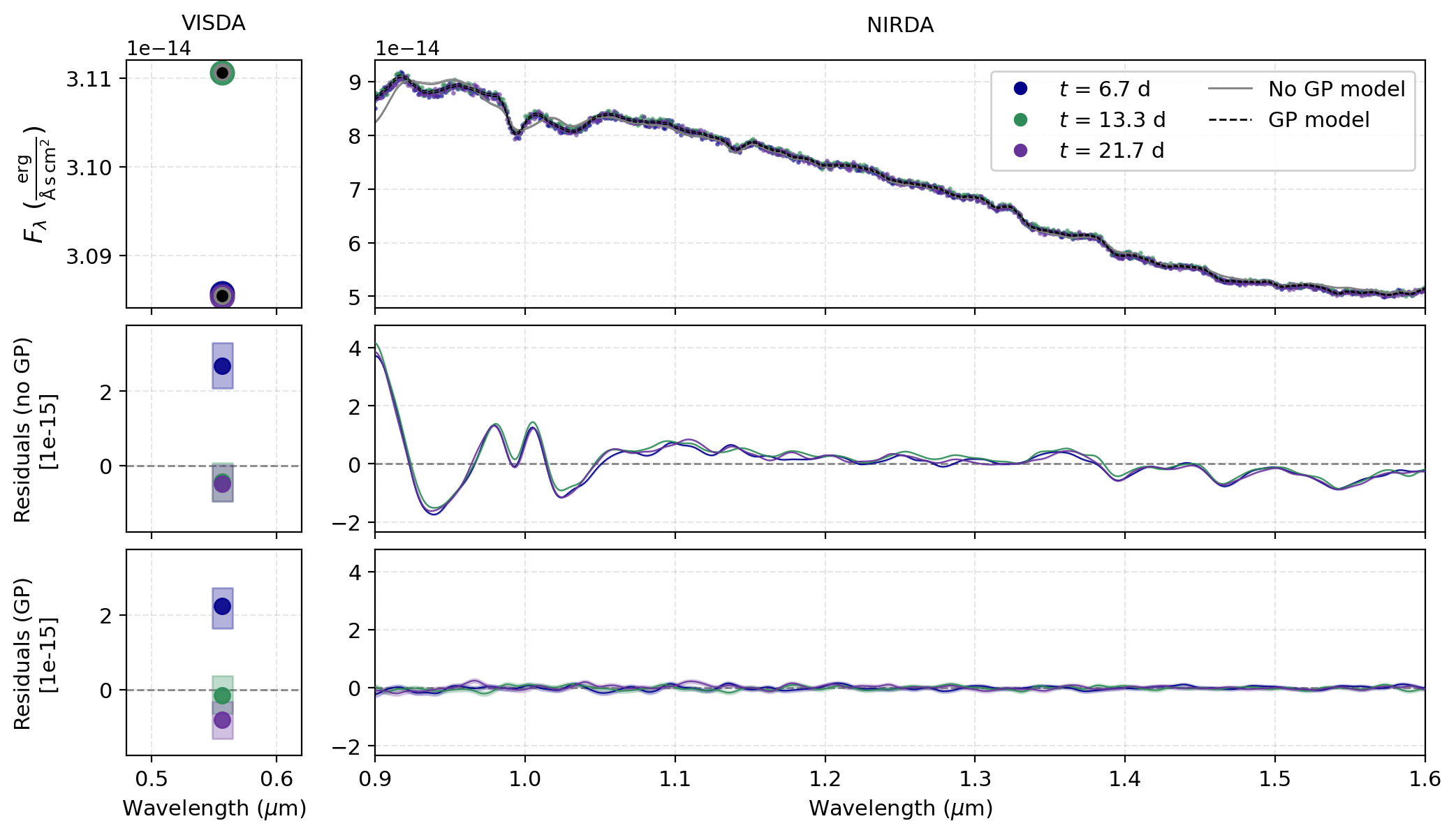}
    \caption{Comparison between the non-GP and GP-aided retrievals for the simulated MFG scenario. Three representative out-of-transit epochs are shown, spanning low, intermediate, and high spot covering fractions. Model uncertainties are shaded in the residual panels. In the NIRDA bandpass, model uncertainties are approximately two orders of magnitude smaller than the plotted scale and are therefore difficult to see.}
    \label{fig:ex_composite_spectra_fits}
\end{figure*}

Figure~\ref{fig:ex_composite_spectra_fits} compares the retrieved disk-integrated spectra from the non-GP and GP-aided retrievals for three representative out-of-transit epochs spanning low, intermediate, and high spot covering fractions. The non-GP retrieval leaves wavelength-dependent residuals, especially toward the shorter wavelengths of the NIRDA band, \vone{whereas the GP-aided retrieval reduces these wavelength-dependent residuals.}

\begin{figure*}
    \centering
    \includegraphics[width=1\linewidth]{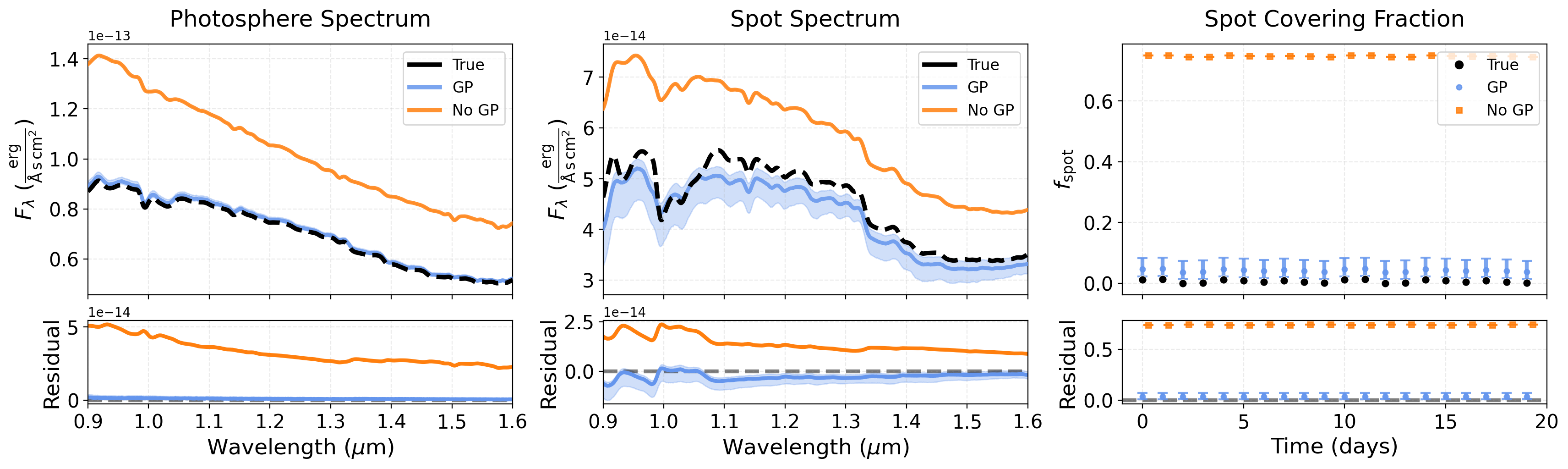}
    \caption{Retrieved latent stellar components for the simulated MFG scenario. Left: photosphere spectrum and residuals. Middle: spot spectrum and residuals. Right: spot covering fraction over time and residuals.}
    \label{fig:ex_component_spectra_fits}
\end{figure*}

Figure~\ref{fig:ex_component_spectra_fits} shows the corresponding inferred latent photosphere spectrum, spot spectrum, and spot covering fraction for the same scenario. \vone{The non-GP retrieval returns stellar components with residuals inconsistent with zero, while the GP-aided retrieval again brings the residuals closer to zero.}

\begin{figure}
    \centering
    \includegraphics[width=\linewidth]{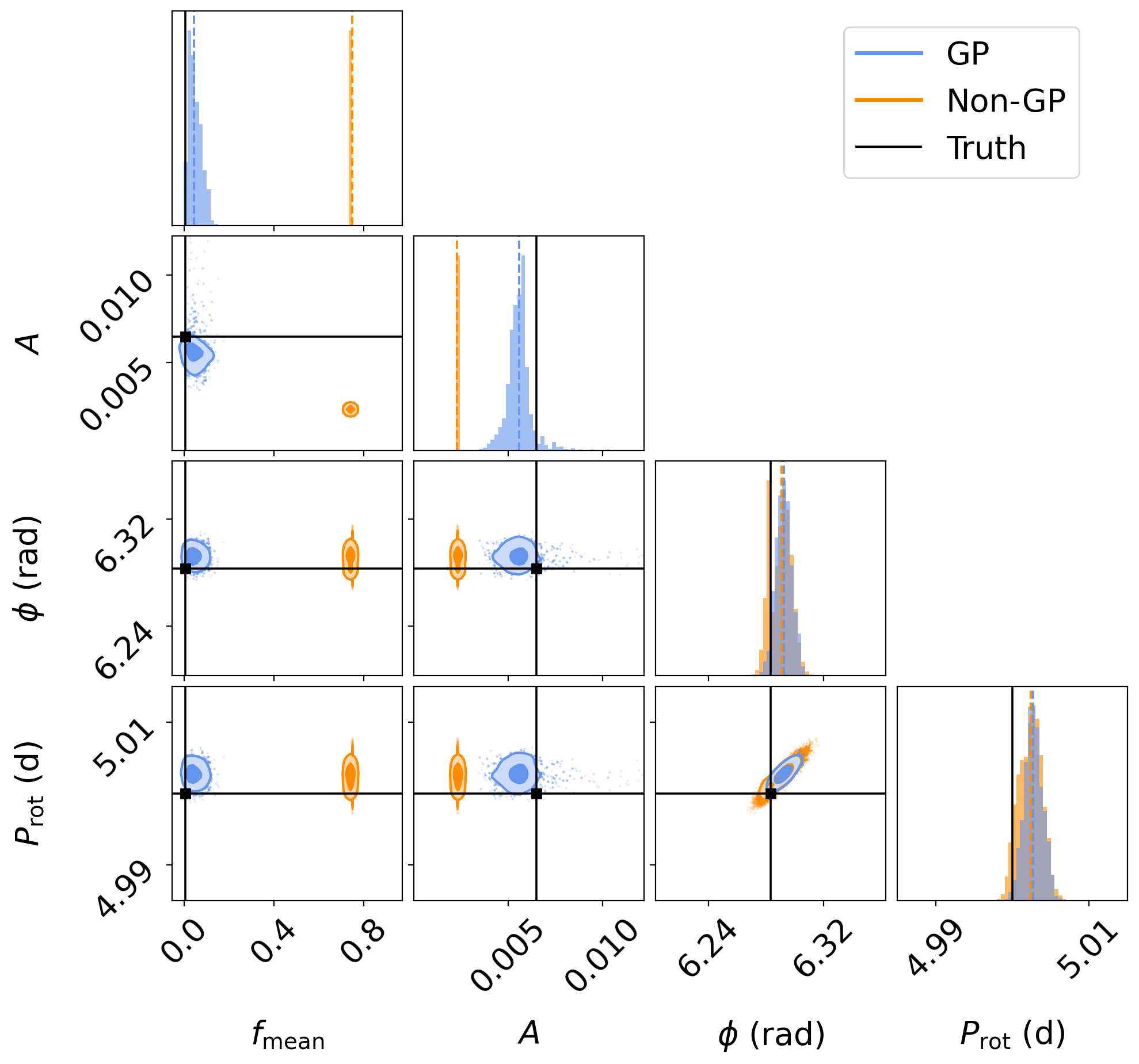}
    \caption{\vone{Posteriors of the spot-covering-fraction parameters for the MFG scenario. Blue and orange show the GP-aided and non-GP retrievals, respectively; black lines mark the injected values.}}
    \label{fig:fspot_corner}
\end{figure}

\vone{Figure~\ref{fig:fspot_corner} shows the corresponding posteriors of the spot-covering-fraction parameters. Both retrievals recover the rotation period and phase, which are largely insensitive to the adopted spectral templates. The GP-aided retrieval yields $f_{\rm mean}$ and $A$ posteriors closer to the injected truth, with inflated uncertainties that reflect the stellar-model mismatch.}

The GP-aided component spectra do not perfectly reproduce the true input spectra at all wavelengths, reflecting the finite information content of the observations, as the component spectra must be inferred through their changing linear combinations rather than being observed independently.

\subsection{Propagation into Atmospheric Inferences}
\label{subsec:results-transmission-correction}

We propagate the stellar retrieval results into the atmospheric retrievals as described in Section~\ref{sec:transmission-propagation} to assess how stellar model treatment affects inferred atmospheric properties. We first present results for the same MFG example, then summarize performance across all scenarios in Section~\ref{subsec:results-all-scenarios}.

\begin{figure}
    \centering
    \includegraphics[width=1\linewidth]{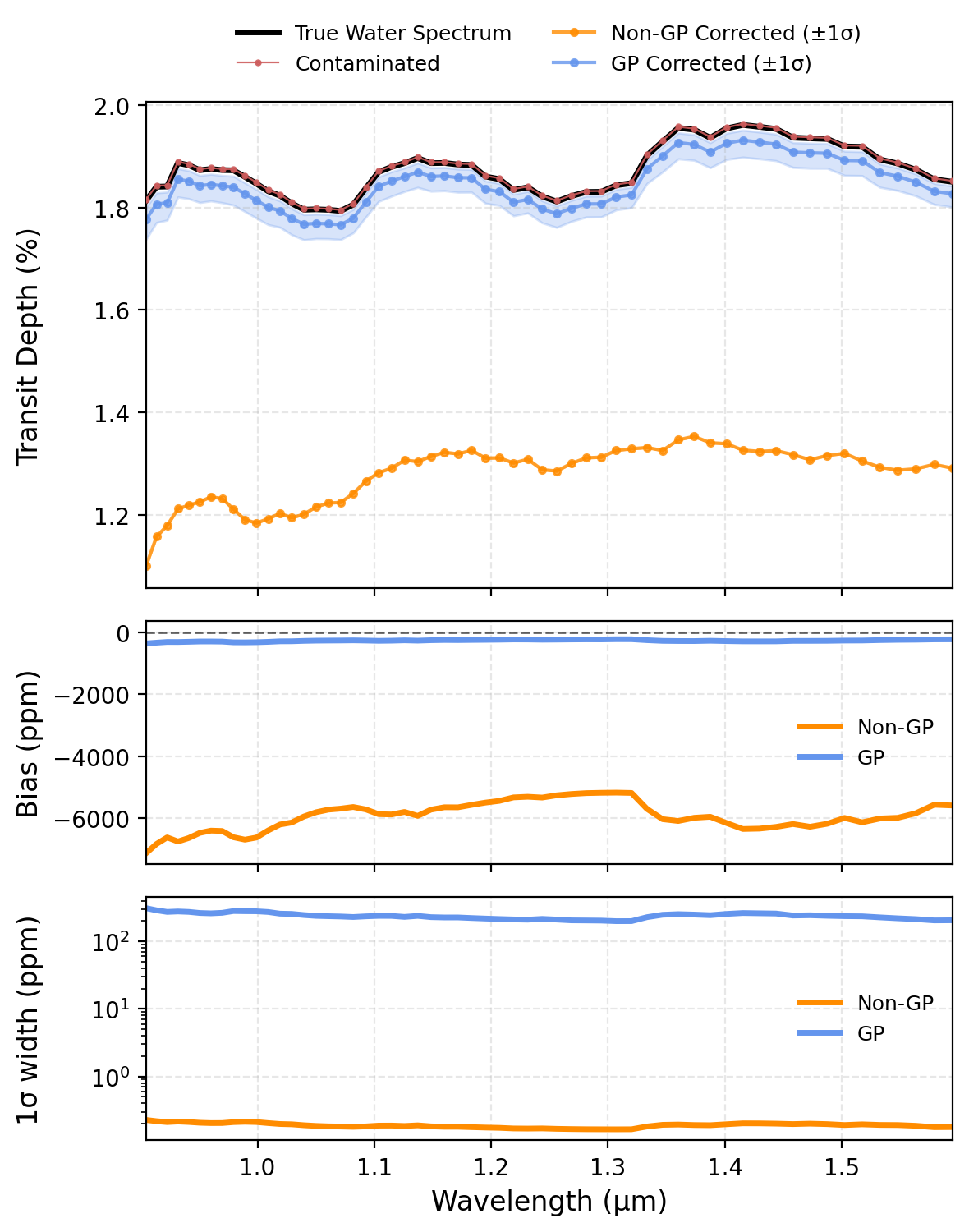}
    \caption{Corrected transmission spectra for the MFG scenario. The top panel compares the true planetary spectrum, the contaminated spectrum, and the median corrected spectra from the non-GP and GP-aided contamination factor posteriors. \vone{Shaded regions show the 16th--84th percentile range propagated from the stellar retrievals; the non-GP range is too narrow to be seen. The middle panel shows the bias of the median corrected spectrum relative to the true spectrum. The bottom shows the width of the corrected spectrum, defined as half the 16th--84th percentile range.}}
    \label{fig:ex_trans_spec_correction}
\end{figure}

Figure~\ref{fig:ex_trans_spec_correction} shows the median corrected transmission spectrum for this scenario. \vone{Because the spot covering fraction in the MFG scenario only spans 0.000--0.013 (Table~\ref{tab:simulated-scenarios}), the stellar contamination is low, and the contaminated transmission spectrum lies close to the injected one. The differences between the non-GP and GP-corrected spectra reflect the effects of the stellar model mismatch. The GP-aided correction reduces the bias and propagates a stellar-model uncertainty.}

\begin{figure}
    \centering
    \includegraphics[width=1\linewidth]{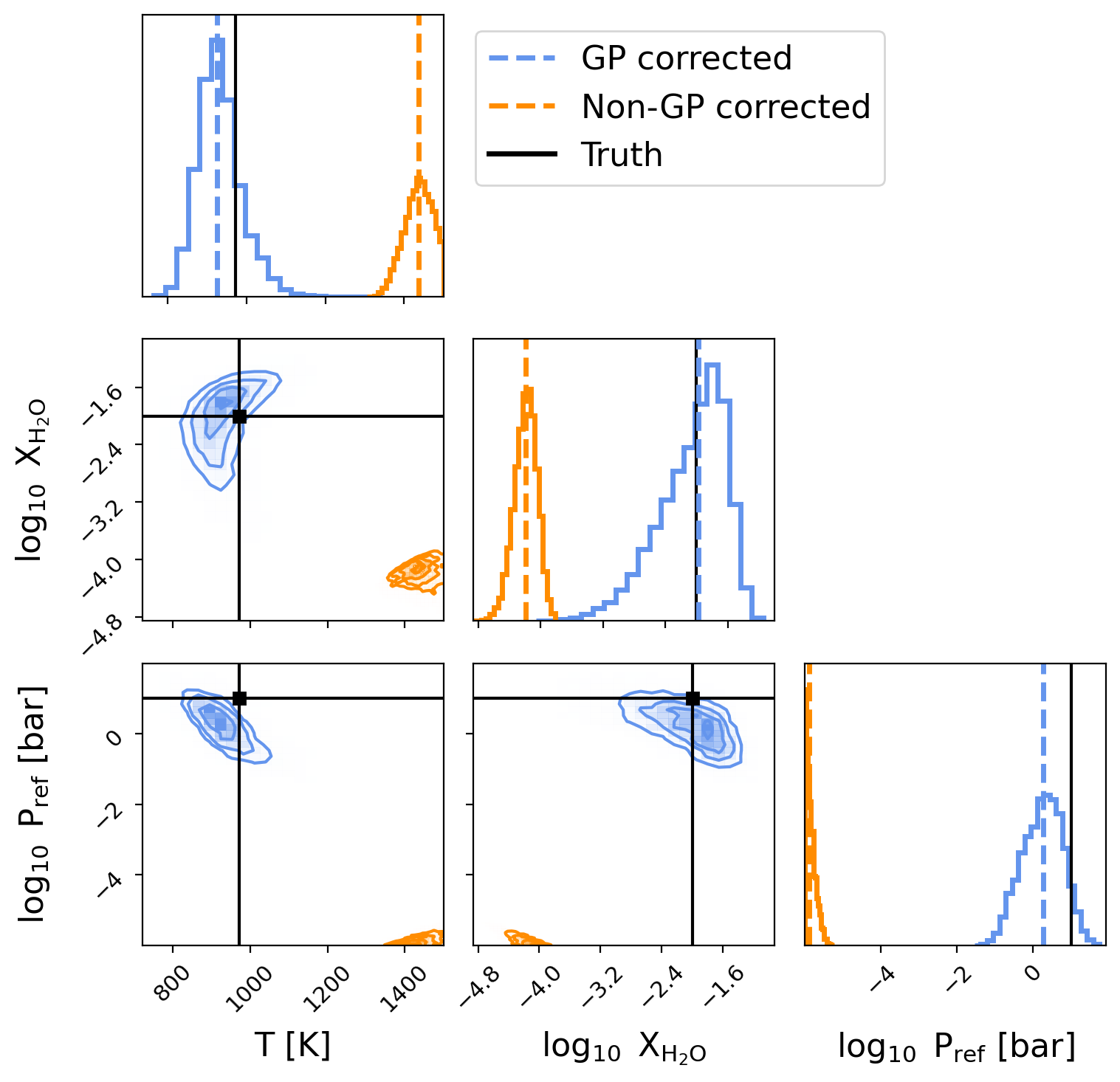}
    \caption{Atmospheric retrieval posteriors for the MFG scenario. Blue and orange show the GP-aided and non-GP cases, respectively; black lines mark the injected atmospheric parameters.}
    \label{fig:ex_atm_retrieval}
\end{figure}

Figure~\ref{fig:ex_atm_retrieval} shows the atmospheric retrieval posteriors for this scenario. The GP-aided correction yields posteriors more consistent with the injected truth.

\subsection{Performance across the full simulated sample}
\label{subsec:results-all-scenarios}

\begin{figure*}
    \centering
    \includegraphics[width=1\linewidth]{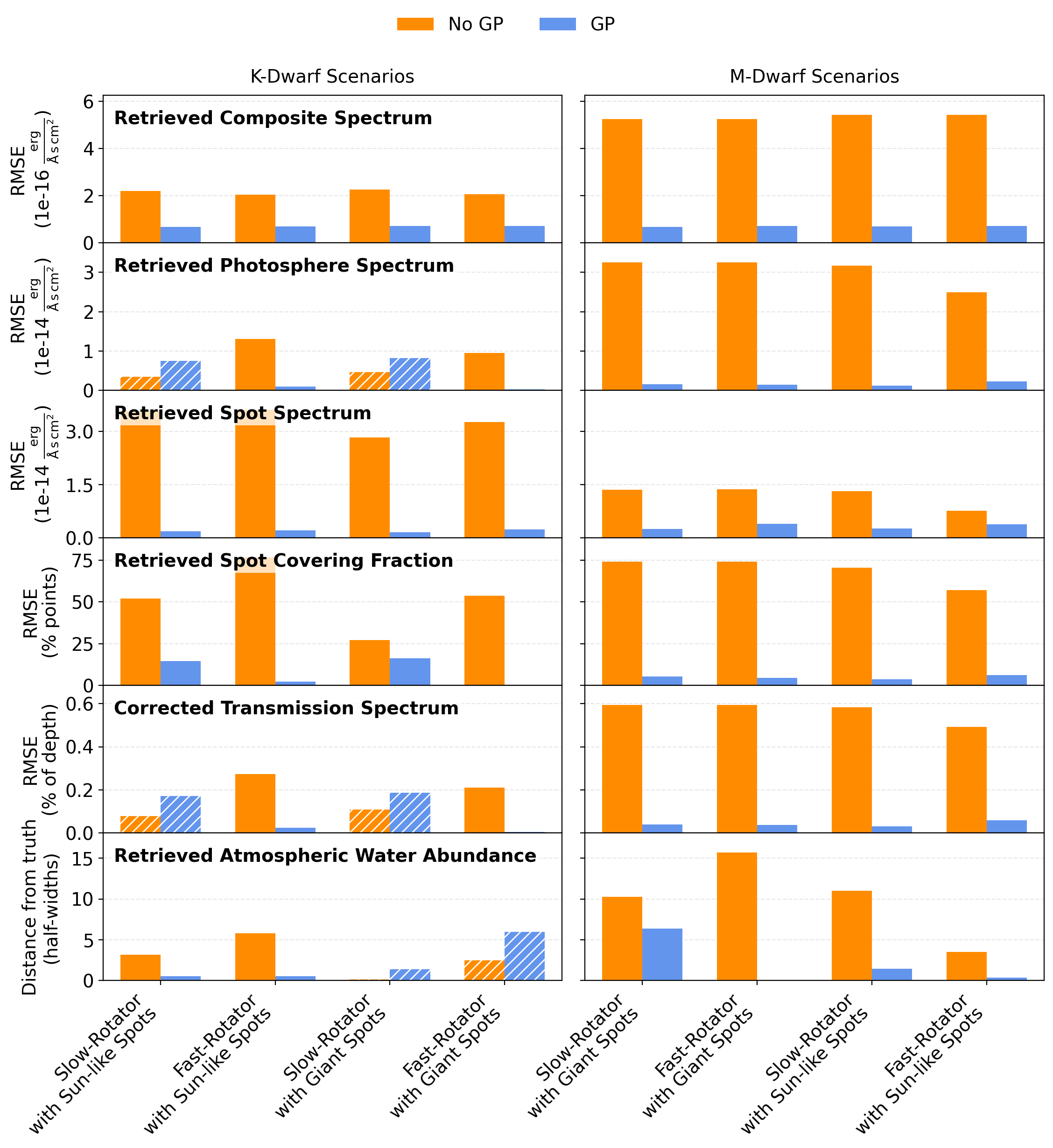}
    \caption{Comparison between the non-GP and GP-aided retrievals across all eight simulated Pandora scenarios, with the K-dwarf scenarios on the left and the M-dwarf scenarios on the right. From top to bottom: RMSE of the recovered composite spectra, photosphere spectra, and spot spectra (in flux units); RMSE of the inferred spot covering fraction \vone{(in absolute percentage points)} and corrected transmission spectrum (in percent \vone{of transit depth}); and the distance of the retrieved $\log X_{\mathrm{H_2O}}$ from truth\vone{, normalized by} the posterior half-width\vone{,} defined as $(p_{84}-p_{16})/2$, to \vone{accommodate} non-Gaussian \vone{posteriors}. The GP RMSE \vone{values} for some cases are too small to be visible on these scales compared with the non-GP cases. Scenarios where the GP-aided retrieval yields a higher RMSE than the non-GP case are indicated with white hatching.}
    \label{fig:all_scenarios}
\end{figure*}

We compare the performance of the non-GP and GP-aided retrievals across all eight simulated Pandora scenarios, summarized in Figure~\ref{fig:all_scenarios}. \vone{The GP-aided retrieval consistently improves the composite spectrum recovery, but this spectrum-level improvement does not guarantee a lower RMSE for every retrieved stellar component and atmospheric parameter. In Figure~\ref{fig:all_scenarios_bias_variance}, we separate each metric into its bias and posterior width, and extend the atmospheric comparison from the water abundance to all three retrieved atmospheric parameters: $\log X_{\mathrm{H_2O}}$, the atmospheric temperature $T$, and the reference pressure $\log P_{\mathrm{ref}}$. }

We also test the case in which the retrieval uses the correct underlying stellar grid, repeating both GP-aided and non-GP retrievals using PHOENIX ACES, the same grid used to generate the simulated data. We present the difference \vone{in} Bayesian evidence as $\Delta \ln Z = \ln Z_{\rm GP} - \ln Z_{\rm non-GP}$ for all scenarios with and without model mismatch in Table~\ref{tab:model_comparison}.

\begin{table}
    \centering
    \caption{Bayesian model comparison between GP and non-GP stellar retrievals. $\Delta \ln Z = \ln Z_{\rm GP} - \ln Z_{\rm non-GP}$, where positive values favor the GP model. \vone{Uncertainties are the nested-sampling uncertainties on $\ln Z_{\rm GP}$ and $\ln Z_{\rm non-GP}$, added in quadrature.} The magnitude of the Bayesian evidence scales with the number of data points; each scenario has 20 time bins $\times$ 702 spectral bins $= 14040$ points. Left column: retrievals with model--data mismatch (PHOENIX ACES data retrieved with PHOENIX NewEra). Right column: retrievals without model--data mismatch (PHOENIX ACES data retrieved with PHOENIX ACES).}
    \label{tab:model_comparison}
    \begin{tabular}{lcc}
    \hline
    \hline
    Scenario & \multicolumn{2}{c}{$\Delta \ln Z$} \\
    \cline{2-3}
     & With mismatch & Without mismatch \\
    \hline
    KSS & $+8591.3 \pm 0.9$  & $-1.4 \pm 1.1$ \\
    KFS & $+7416.4 \pm 0.9$  & $-0.9 \pm 1.2$ \\
    KSG & $+8733.5 \pm 1.2$  & $-2.5 \pm 1.1$ \\
    KFG & $+7368.1 \pm 1.0$  & $+1.1 \pm 1.5$ \\
    MSS & $+28020.6 \pm 1.2$ & $+0.2 \pm 0.7$ \\
    MFS & $+29926.3 \pm 1.0$ & $+0.6 \pm 1.2$ \\
    MSG & $+28017.9 \pm 1.1$ & $+2.1 \pm 1.0$ \\
    MFG & $+27652.9 \pm 1.2$ & $+0.6 \pm 0.9$ \\
    \hline
    \end{tabular}
\end{table}

\section{Discussion}\label{sec:discussion}

\subsection{Improving stellar and atmospheric retrievals}\label{subsec:improving_retrievals_withgps}

\vone{The GP-aided framework consistently yields a higher Bayesian evidence in the face of model mismatch (Table~\ref{tab:model_comparison}) and improves the recovered composite spectra across all scenarios (Figure~\ref{fig:all_scenarios}).} The non-GP retrieval is limited to the adopted PHOENIX NewEra grid and therefore cannot fully reproduce the PHOENIX ACES-generated data when the two grids differ. The GP-aided retrieval absorbs the wavelength-dependent residuals into the latent component spectra, \vone{reducing biases in the recovered composite stellar spectra.}

\begin{figure}
    \centering
    \includegraphics[width=\linewidth]{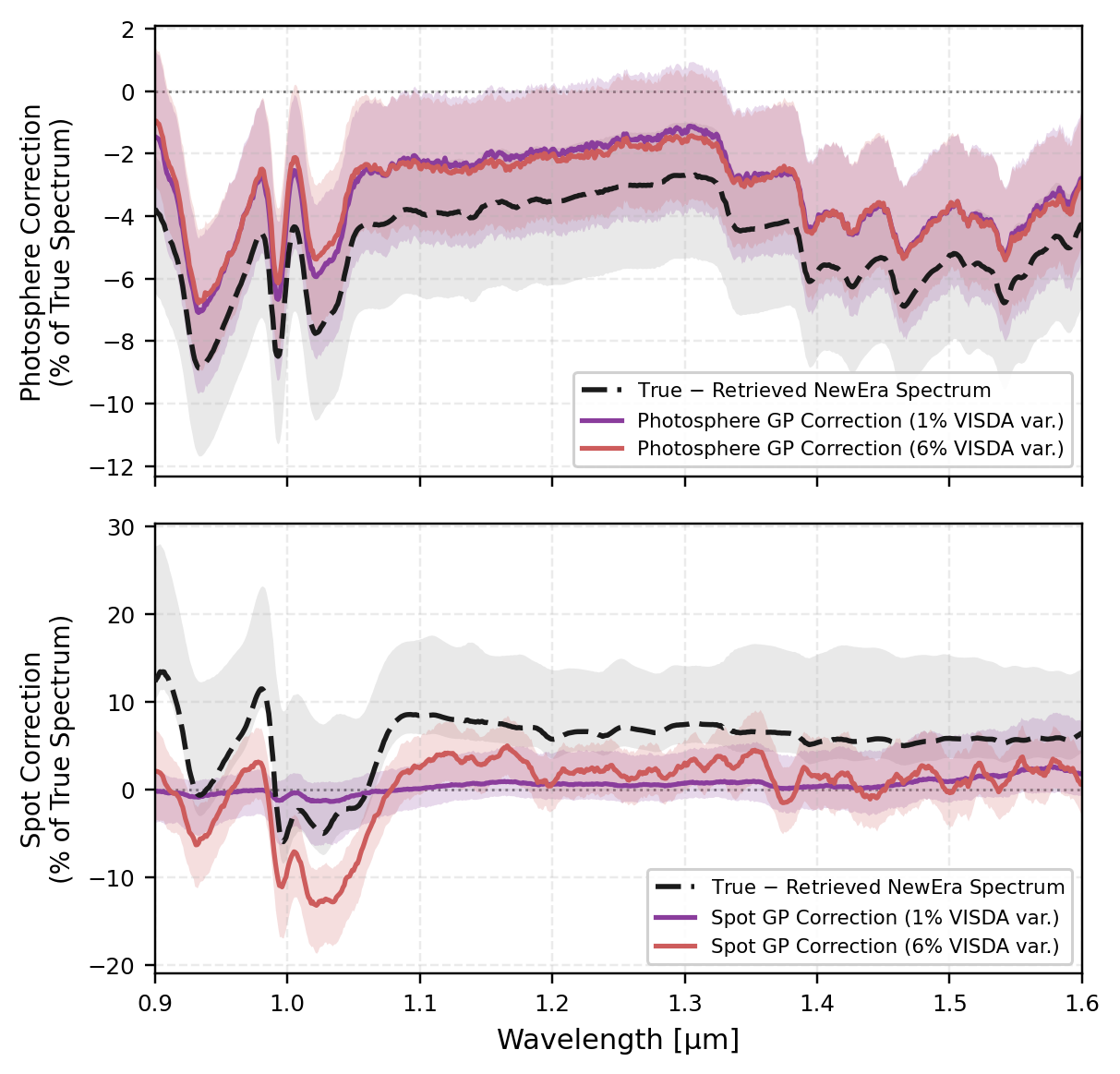}
    \caption{\vone{GP spectral contributions to the photosphere (top) and spot (bottom) components in the MFG scenario, in percent of the injected PHOENIX ACES component spectrum. Purple curves show the GP correction for the MFG scenario in our synthetic dataset, which has 1\% peak-to-peak VISDA variability. Pink shows the GP corrections for the original MFG dataset from \citet{rackham2026pand}, with roughly 6\% peak-to-peak VISDA variability. Colored curves and bands show the median and 16th--84th percentile regions. Black dashed curves and gray bands show the difference between the injected spectrum and the PHOENIX NewEra mean spectrum evaluated at the retrieved stellar parameters, which is the correction that would recover the injected spectrum exactly.}}
    \label{fig:gp_mismatch}
\end{figure}

\vone{The recovery of individual stellar components is not always improved with the GP-aided retrieval, as they are less directly constrained by observations. As shown in Section~\ref{subsec:baseline}, under the adopted sinusoidal filling-factor model, observations constrain the mean composite spectrum and rotational modulation more directly than the individual photosphere spectrum, spot spectrum, and spot covering fractions, allowing for degenerate component solutions. Disentangling the photosphere and spot spectra also relies on sufficient variation in the spot covering fraction, since the spot spectrum is constrained mainly through the rotational modulation it produces. In our simulated Pandora observations, with 1\% peak-to-peak VISDA variability, the spot contributes little to the observed flux, which limits how well the GP can identify the spot model--data mismatch. Figure~\ref{fig:gp_mismatch} presents the photosphere and spot GP corrections for the MFG scenario. We show two cases: the simulated MFG observations used throughout this work, with 1\% peak-to-peak VISDA variability, and the original MFG dataset from \citet{rackham2026pand}, available on Zenodo\footnote{DOI: \dataset[10.5281/zenodo.18624584]{\doi{10.5281/zenodo.18624584}}}, which has a larger spot covering-fraction range ($\sim$0.0006--0.08) and approximately 6\% peak-to-peak VISDA variability.}

\vone{With 1\% VISDA variability, the photosphere GP correction recovers the main wavelength structure of the photosphere mismatch, while the spot GP correction stays close to zero and does not reproduce the spot mismatch. The spot covering fraction in this scenario is only $\sim$0.0001--0.013 (Table~\ref{tab:simulated-scenarios}), so the spot spectrum contributes little of the observed flux. With 6\% variability, the stronger spot signal allows the spot GP correction to follow the wavelength structure of the spot mismatch, while the photosphere correction is largely unchanged. The ability of the GP to effectively capture the spot mismatch therefore depends on the amplitude of the spot-driven variability.}

\vone{A weakly constrained spot GP does not necessarily lead to a worse retrieved spot spectrum. The non-GP retrieval can absorb the photosphere mismatch only by shifting the stellar parameters, and the spot parameters can absorb remaining mismatch. The non-GP retrievals therefore lock into tightly constrained but biased solutions. In the GP-aided retrieval, the photosphere GP absorbs the photosphere mismatch, which can also pull the spot spectrum toward the truth. In most scenarios, the GP-aided retrieval recovers a more accurate photosphere spectrum, spot spectrum, and spot covering fraction compared to the non-GP retrieval (Figure~\ref{fig:all_scenarios}).}

\textbf{The contamination factor used to correct the transmission spectrum depends explicitly on the inferred photosphere spectrum (Equation~\ref{eq:contamination-factor}). In the KSG scenario, the GP-aided retrieval yields a higher RMSE on the retrieved photosphere spectrum, which translates to a higher RMSE on the corrected transmission spectrum and retrieved water abundance, highlighting how this intrinsic degeneracy between stellar components can affect retrieved atmospheric properties. However, performance in stellar retrievals does not directly translate to performance in atmospheric retrievals due to degeneracies between atmospheric parameters: the retrieved H$_2$O abundance, atmospheric temperature, and reference pressure can combine degenerately to reproduce changes in the offset and wavelength-dependent shape of the spectrum \citep[e.g.][]{heng2017, welbanks2019}. In the KFG scenario, the GP-aided retrieval yields a larger distance from truth on the retrieved water abundance, despite lower RMSE on the retrieved photosphere spectrum, while in the KSS scneario, the GP-aided retrieval yields a smaller distance from truth on the retrieved water abundance, despite higher RMSE on the retrieved photosphere spectrum.}

\vone{Across all M-dwarf scenarios, the GP-aided corrections yields a consistent improvement over the non-GP corrections, as the ACES-NewEra mismatch is larger for the M-dwarf scenarios compared to the K-dwarf scenarios.} The GP-aided approach is therefore most valuable when stellar model mismatch is large enough to limit deterministic contamination corrections. 

\vone{Figure~\ref{fig:all_scenarios_bias_variance} shows the bias-variance decomposition of Figure~\ref{fig:all_scenarios}. Because the non-GP stellar retrieval has no explicit model-discrepancy term, it under-propagates the uncertainty in the stellar models. The GP-aided retrieval instead marginalizes over wavelength-correlated stellar-model uncertainty, which reduces the bias at the cost of a larger but more realistic variance. The GP-aided framework therefore improves inferences by replacing overconfident and potentially biased constraints with broader but less biased posteriors. We note that the retrieved water abundance under the KSG and MSG scenarios and the retrieved reference pressure under the KSG scenario have higher variance under the non-GP correction. These broader posteriors do not reflect propagated stellar uncertainty; they arise because the biased contamination correction forces the atmospheric model toward the edge of the temperature prior and inflates the degeneracy among atmospheric parameters.}

\subsection{Inferring the source of model--data mismatch}
\label{subsec:mismatch_source}

While GPs offer empirically-learned corrections rather than physics-driven corrections, their inferred wavelength structure can still provide insight into the missing physics that drive the mismatch, as shown in Figure~\ref{fig:gp_mismatch}. \vone{For our simulated dataset with 1\% peak-to-peak VISDA variability, the photosphere GP component recovers the main wavelength structure of the underlying model mismatch, and the spot GP component can recover the wavelength structure of the mismatch when the spot contribution is increased.}

\vone{The vertical offsets of the GP contributions reflect the degeneracy discussed in Section~\ref{subsec:improving_retrievals_withgps}. Deterministic stellar parameters are inferred simultaneously with the GP hyperparameters, so changes in $T_{\rm phot}$, $T_{\rm spot}$, and $f_{\rm spot}$ can absorb some of the same wavelength-dependent structure captured by the GP. This is especially relevant near molecular bandheads, where opacity differences between model grids and temperature or spot-filling-factor changes can produce similar spectral structure.}

\vone{Nonetheless, the structure of the GP contribution can follow that of the mismatch, particularly below $\sim 1.1~\mu$m, where the ACES-NewEra discrepancy is largest.} This wavelength region is dominated by FeH and TiO opacity, molecules whose line lists differ between the two grids: NewEra adopts updated ExoMol data for TiO \citep{mckemmish2020} and FeH \citep{dulick2003, burrows2005, bernath2020}, whereas ACES relies on earlier molecular databases \citep{phoenix}.

\vone{The GP corrections can therefore identify spectral regions where the stellar models are least adequate, but only for components that contribute enough of the observed flux to be constrained. Unlike direct residual inspection, the GP model quantifies the amplitude and wavelength scale of the discrepancy and infers the level of correlated model-error structure supported by the data given the observational uncertainties. In this way, the GP-aided framework can serve both as a tool for marginalizing over stellar-model systematics and as a guide to where stellar models, such as their molecular line lists, most need improvement.}

\subsection{When to use the GP-aided framework}\label{subsec:when_to_use}
For a real observation, the true extent of model--data mismatch is unknown a priori. Therefore, it is important to establish a metric to determine when GP components are warranted. Table~\ref{tab:model_comparison} demonstrates that Bayesian model comparison can serve as this guideline, as Bayesian evidence rewards goodness of fit while penalizing model complexity. Thus, the GP-aided model is penalized unless the improvement in the likelihood is sufficient to justify the added complexity.

\vone{When the retrieval grid matches the grid used to generate the data, the GP-aided retrieval effectively reduces to the non-GP retrieval. The posterior of the photosphere GP amplitude collapses onto its lower prior bound, and its length scale becomes uninformative, since a correlation length has no effect on the likelihood once the GP amplitude is negligible. The spot GP hyperparameters remain uninformative because the spot contributes only a small fraction of the total flux (Section~\ref{subsec:improving_retrievals_withgps}). With the GP terms suppressed, the GP-aided fit performs comparably to the non-GP fit, and the resulting $\Delta \ln Z$ values are consistent with zero within their uncertainties (Table~\ref{tab:model_comparison}). In contrast, when the retrieval grid differs from the generating grid, the photosphere GP amplitude is constrained above the lower prior bound, and $\Delta \ln Z$ reaches $\sim 10^{4}$ in favor of the GP-aided model.}

For real observations, it is \vone{therefore} important to compare both non-GP and GP-aided retrievals to determine whether the GP corrections are justified. A \vone{large} positive $\Delta \ln Z = \ln Z_{\rm GP} - \ln Z_{\rm non-GP}$ \vone{along with a GP amplitude constrained away from its lower prior bound} indicates that the adopted stellar grid does not adequately capture the observed spectra and that the GP-aided framework should be used for contamination correction. Conversely, a negative or near-zero $\Delta \ln Z$ suggests that the deterministic stellar model is sufficient and the added GP flexibility is unwarranted. 

Additional diagnostics \vone{should} be used as complementary checks, including posterior predictive checks, cross-validation tests, and inspection of wavelength-correlated residuals. \vone{Additionally, a positive $\Delta \ln Z$ indicates only that the more flexible model describes the data better; it does not by itself identify the physical origin of the discrepancy, which could also arise from instrumental systematics or additional surface components. These possibilities should therefore be considered before the GP components are interpreted as a correction to the stellar atmosphere models alone.}

\subsection{Future extensions}\label{subsec:future_extensions}

The GP-aided stellar-contamination retrieval framework presented here is designed to test whether Pandora's time-variable spectra can empirically correct wavelength-dependent stellar model mismatch. To isolate this problem, we make simplifying assumptions about the origin of the stellar-model mismatch, the time evolution of the spot covering fraction, and the complexity of the stellar surface. We summarize these in Table~\ref{tab:future_extensions} and highlight future extensions that can be implemented.

\begin{table*}
\centering
\caption{Summary of simplifying assumptions adopted in the GP-aided stellar retrieval framework, their justification in this work, and possible future extensions.}
\label{tab:future_extensions}
\scriptsize
\begin{tabular}{lll}
\hline
\textbf{Simplifying assumption} & \textbf{Justification} & \textbf{Future extension} \\
\hline
\\

\begin{minipage}[t]{0.28\textwidth}
We adopt \vone{a revised} PHOENIX ACES-generated \vone{dataset based on} \citet{rackham2026pand}\vone{, with roughly 1\% VISDA variability,} as truth and run retrievals with PHOENIX NewEra as a proxy for real model--data mismatch.
\end{minipage}
&
\begin{minipage}[t]{0.33\textwidth}
This provides a controlled test in which the true component spectra and spot covering fractions are known, allowing us to quantify the GP-aided framework performance. Additionally, the simulated model-model mismatch mimics real model--data mismatch behavior.
\end{minipage}
&
\begin{minipage}[t]{0.33\textwidth}
Rebuild the simulated dataset with different stellar model grids and test different grids during retrieval to characterize GP performance across a broader range of mismatch.
\end{minipage}
\\
\\
\hline
\\
\begin{minipage}[t]{0.28\textwidth}
The spot covering fraction is assumed to follow a stable sinusoid. Spot sizes, temperatures, and \vone{patterns} do not evolve over the observing baseline, and the retrieval adopts the same sinusoidal functional form \vone{used to generate the data}.
\end{minipage}
&
\begin{minipage}[t]{0.33\textwidth}
This isolates the impact of the wavelength-domain stellar spectral mismatch without introducing additional uncertainty from a complex time-domain variability model.
\end{minipage}
&
\begin{minipage}[t]{0.33\textwidth}
Incorporate more \vone{flexible and realistic} time-domain models for stellar variability and spot evolution, such as quasi-periodic GPs or physically motivated stellar variability models \vone{to prevent biasing inferred stellar components from a biased temporal model}.
\end{minipage}
\\
\\
\hline
\\
\begin{minipage}[t]{0.28\textwidth}
The simulated visits for each scenario are evenly distributed over approximately 30 days and are assumed to provide sufficient phase coverage of the stellar activity cycle.
\end{minipage}
&
\begin{minipage}[t]{0.33\textwidth}
This provides a controlled sampling of different spot covering fractions and maximizes the information available to disentangle the latent photosphere and spot spectra.
\end{minipage}
&
\begin{minipage}[t]{0.33\textwidth}
Assess how realistic Pandora scheduling, including uneven sampling and aliasing with the stellar rotation period, affects recovery of the latent stellar components.
\end{minipage}
\\
\\
\hline
\\
\begin{minipage}[t]{0.28\textwidth}
The stellar surface is modeled with only two components: a quiet photosphere and cool spots.
\end{minipage}
&
\begin{minipage}[t]{0.33\textwidth}
This provides a controlled test of whether Pandora's time-variable observations can disentangle two latent spectral components.
\end{minipage}
&
\begin{minipage}[t]{0.33\textwidth}
Extend the framework to include additional stellar components, such as faculae, plages, multiple spot-temperature populations, or continuous temperature distributions.
\end{minipage}
\\
\\
\hline
\\
\begin{minipage}[t]{0.28\textwidth} 
\vone{The planet is assumed to transit an immaculate photosphere.}

\end{minipage}
&
\begin{minipage}[t]{0.33\textwidth}
This isolates the effect of wavelength-domain stellar-model mismatch and matches the standard unocculted-spot contamination prescription used for the atmospheric retrieval tests.
\end{minipage}
&
\begin{minipage}[t]{0.33\textwidth}
\vone{Extend the framework to account for heterogeneous transit chords, including spot- and facula-crossing events \citep[e.g.,][]{Espinoza2019}, and the dependence of stellar contamination on active-region latitude and planetary impact parameter \citep[e.g.,][]{Glidden2026}.}
\end{minipage}
\\
\\
\hline
\\
\begin{minipage}[t]{0.28\textwidth}
\vone{Photosphere and spot GP corrections are assumed to be independent, though inaccurate molecular opacities could produce correlated structure in both component spectra.}
\end{minipage}
&
\begin{minipage}[t]{0.33\textwidth}
This keeps the covariance structure tractable. 
\end{minipage}
&
\begin{minipage}[t]{0.33\textwidth}
Use a shared latent GP component to model wavelength-dependent stellar-model discrepancies that are correlated between the photosphere and spot spectra.
\end{minipage}
\\
\\
\end{tabular}
\end{table*}

Despite these limitations, this work provides a framework for using GPs to disentangle time-varying spectroscopic observations and empirically infer stellar-contamination corrections. The framework presented here is modular: the stellar grids are used as the GP mean functions, the time-domain model is used for the filling factors, and the number of stellar surface components can be changed. Therefore, the limitations presented in Table~\ref{tab:future_extensions} can be addressed independently and incorporated directly into the existing retrieval framework.

\section{Conclusion}\label{sec:conclusion}

\vone{In this work, we present a GP-aided framework that leverages Pandora's long-baseline observations to empirically learn the stellar spectra needed for stellar-contamination corrections. Rather than treating stellar atmosphere grids as exact spectral templates, we use them as physically motivated mean functions and allow the data to learn wavelength-dependent deviations from those models through GPs. One GP describes deviations from the photosphere model, while a second GP describes deviations from the spot model. This allows the retrieval to absorb model--data mismatch while retaining the physical structure of a two-component stellar-contamination model.}

\vone{Our framework directly propagates the empirically learned stellar model into the atmospheric retrieval, allowing the atmospheric inference to marginalize over the stellar-contamination uncertainty, including wavelength-correlated uncertainty from imperfect stellar models. This differs from conventional corrections that rely on deterministic and potentially biased stellar models, which can yield biased atmospheric constraints.}

We test this framework using simulated Pandora observations for eight representative K- and M-dwarf stellar activity scenarios. Across the full simulated sample, the GP-aided retrieval consistently improves the recovery of the disk-integrated composite stellar spectra relative to the deterministic non-GP retrieval. \vone{This spectrum-level improvement can reduce model-driven biases in inferred atmospheric properties, although it does not guarantee improved recovery of atmospheric parameters. Each observation directly constrains the composite spectrum, but the photosphere spectrum, spot spectrum, and spot covering fraction must be inferred, and degeneracies among these components can bias the corrected transmission spectrum. Additionally, degeneracies} among $T$, $P_{\rm ref}$, and H$_2$O abundance, together with sensitivity to wavelength-dependent residual structure, can cause small differences in the corrected spectra to be absorbed by different combinations of atmospheric parameters. The GP-induced benefits are most apparent for the M-dwarf scenarios, where the model--data mismatch between stellar grids is largest.

\vone{These results demonstrate that our GP-aided spectral disentangling method is a promising approach \vone{for leveraging} time-variable observations to empirically learn the stellar spectra required for contamination corrections, though future extensions that broaden the tested stellar-model mismatch, relax the simplified time-domain and scheduling assumptions, and include additional stellar surface components will further improve the framework's applicability. More broadly, similar GP-aided approaches can be applied to future transmission-spectroscopy datasets from missions such as Ariel \citep{tinetti2018} and Nautilus \citep{apai2019}, where stellar contamination will remain a key systematic in the atmospheric characterization of planets orbiting active stars.}

\begin{acknowledgments}

We thank the anonymous referee for their insightful comments, which have greatly improved the manuscript. This work was performed under the auspices of the U.S. Department of Energy by Lawrence Livermore National Laboratory under Contract DE-AC52-07NA27344. 
The document number is LLNL-JRNL-2023944. Z.K. and P.M. were supported by the LLNL Space Science Institute and the LLNL Academic Engagement Office. Pandora is supported by NASA’s Astrophysics Pioneers Program. This material is based upon work partly supported by the National Aeronautics and Space Administration under Agreement No. 80NSSC21K0593 for the program ``Alien Earths.'' The results reported herein benefited from collaborations and/or information exchange within NASA’s Nexus for Exoplanet System Science (NExSS) research coordination network sponsored by NASA’s Science Mission Directorate. T.O.F. would like to acknowledge support from the NASA Postdoctoral Program by an appointment at the NASA Goddard Space Flight Center, administered by Oak Ridge Associated Universities under contract with NASA. B.J.H. is supported by an appointment to the NASA Postdoctoral Program at NASA Goddard Space Flight Center, administered by Oak Ridge Associated Universities under contract with NASA (ORAU-80HQTR21CA005).

\end{acknowledgments}

\software{NumPy \citep{harris_array_2020}, AstroPy \citep{the_astropy_collaboration_astropy_2022}, Matplotlib \citep{hunter_matplotlib_2007}, corner \citep{corner}, MultiNest \citep{feroz_multinest_2009} via PyMultiNest \citep{buchner_x-ray_2014}, dynesty \citep{speagle2020dynesty}, UltraNest \citep{Buchner2016, Buchner2019}, POSEIDON \citep{macdonald2017, macdonald2023poseidon}, pandora-sat and pandora-sim \citep{hedges2024}, speclib \citep{rackham2024, speclib_v0.1.0b11}}

\appendix
\section{Bias and Variance of Retrievals}
\label{app:bias_variance}

\begin{figure*}
    \centering
    \includegraphics[width=1\textwidth]{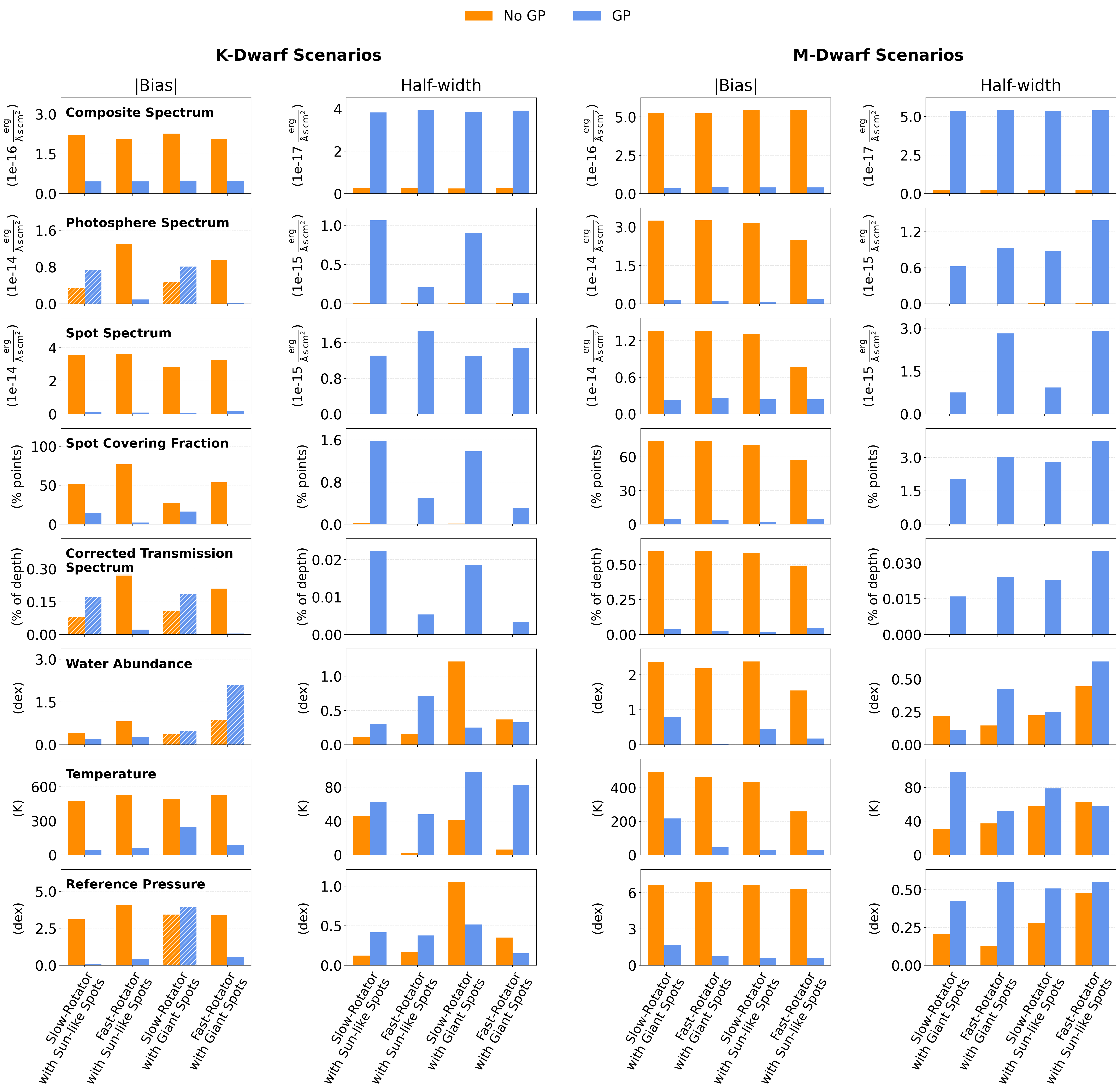}
    \caption{Bias and posterior width of the non-GP and GP-aided retrievals across all eight simulated Pandora scenarios, expanded from Figure~\ref{fig:all_scenarios}. The left two columns show the K-dwarf scenarios and the right two columns show the M-dwarf scenarios. Within each pair, the first column shows the mean absolute bias, $|\mathrm{median} - \mathrm{truth}|$, and the second shows the mean posterior half-width, $w = (p_{84} - p_{16})/2$. From top to bottom: the recovered composite, photosphere, and spot spectra (in flux units, averaged over wavelength); the inferred spot covering fraction (in absolute percentage points, averaged over the 20 epochs); the corrected transmission spectrum (in percent of the transit depth averaged over wavelength); and the retrieved water abundance (in dex), temperature (in K), and reference pressure (in dex). The non-GP half-widths for the photosphere spectrum, spot spectrum, spot covering fraction, and corrected transmission spectrum are too small to be visible on these scales compared with the GP-aided half-widths. Scenarios where the GP-aided bias exceeds the non-GP bias are indicated with white hatching.}
    \label{fig:all_scenarios_bias_variance}
\end{figure*}

\bibliography{refs}{}
\bibliographystyle{aasjournalv7}

\end{document}